\let\ifluatex\relax
\documentclass[sigconf]{acmart}

\acmConference[SIGMOD '23]{Proceedings of the 2023 International Conference on Management of Data}{June 18-23, 2022}{Seattle, USA}
\acmYear{2023}
\acmISBN{} 
\acmDOI{} 
\startPage{1}

\usepackage{booktabs}   
\usepackage{subcaption} 
\usepackage{caption}
\usepackage{colortbl}
\usepackage{style}
\usepackage{bigstrut}
\usepackage{microtype}
\usepackage{tabularx}
\usepackage{multirow}
\usepackage{multicol}
\usepackage{rotating}
\usepackage{lipsum}
\usepackage{balance}
\usepackage{mfirstuc}
\usepackage{titlecaps}
\usepackage{listings}
\usepackage{float}
\usepackage{tcolorbox}
\usepackage[normalem]{ulem}
\usepackage[rightcaption]{sidecap}
\usepackage{titlecaps}
\usepackage[compact]{titlesec}
 \titlespacing*{\section}
 {0pt}{.7ex plus .5ex minus .2ex}{.3ex plus .1ex}
 \titlespacing*{\subsection}
 {0pt}{.5ex plus .5ex minus .2ex}{.3ex plus .1ex}

\usepackage[font=small,labelfont=bf,skip=2pt]{caption}
\usepackage{cleveref}

\usepackage{soul}
\definecolor{lightgray}{rgb}{0.94,0.94,0.94}
\sethlcolor{lightgray}

  \newcommand{\fname}[1]{\textsc{#1}}
  \newcommand{\vname}[1]{\mathit{#1}}

  \newcommand{\newchange}[1]{{#1}}
  \newcommand{\revision}[1]{{\color{black}#1}}

  \newcommand{\hashbag}{hash bag\xspace}
  \newcommand{\lelist}{LE-list\xspace}
  \newcommand{\tail}{\vname{tail}\xspace}
  \newcommand{\sample}{\vname{sample}}

  \newcommand{\bagarray}{\vname{bag}\xspace}
  \newcommand{\bagput}{\fname{Insert}\xspace}
  \newcommand{\bagforall}{\fname{ForAll}\xspace}
  \newcommand{\bagpack}{\fname{ExtractAll}\xspace}
  \newcommand{\bagfind}{\fname{Find}\xspace}
  \newcommand{\bagresize}{\fname{try\_resize}\xspace}
  \newcommand{\knn}{$k$-NN\xspace}
  
  \newcommand{\ff}{\mathcal{F}}
  \newcommand{\vgc}{vertical granularity control}
  \newcommand{\VGC}{VGC}

  \newcommand{\implementationname}[1]{\textsf{#1}}
  \newcommand{\connectit}{\implementationname{ConnectIt}\xspace}
  \newcommand{\ispan}{\implementationname{iSpan}\xspace}
  \newcommand{\multistep}{\implementationname{Multi-step}\xspace}
  
  \newcommand{\gbbs}{\implementationname{GBBS}\xspace}
  \newcommand{\seq}{\implementationname{SEQ}\xspace}
  \newcommand{\parlay}{\implementationname{ParlayLib}\xspace}

  \newcommand{\frontier}{\mathcal{F}}

\newcommand{\sccone}{\mathit{SCC}_1}

  \newcommand{\modelop}[1]{\texttt{#1}}
  \newcommand{\forkins}{\modelop{fork}}

  \newcommand{\thread}{thread}

  \newcommand{\faa}{{\texttt{fetch\_and\_add}}}
  
  \newcommand{\cas}{{\texttt{compare\_and\_swap}}}
  \newcommand{\CAS}{{\texttt{CAS}}}

  \newcommand{\true}{\emph{true}\xspace}
  \newcommand{\false}{\emph{false}\xspace}

  \newcommand{\ifconference}{{{\ifx\fullversion\undefined}}}

\makeatletter
\def\dfnt@space@setup{%
  \dfnt@preskip=\parskip
    \dfnt@postskip=0pt}
\makeatother

\newtheoremstyle{exampstyle}
  {.05in} 
  {.05in} 
  {} 
  {.5em} 
  {\sc \bfseries} 
  {.} 
  {.5em} 
  {} 
\theoremstyle{exampstyle} 
\theoremstyle{exampstyle} 

\makeatletter
\renewenvironment{proof}[1][\proofname]{\par
  \vspace{-\topsep}
  \pushQED{\qed}%
  \normalfont
  \topsep0pt \partopsep0pt 
  \trivlist
  \item[\hskip\labelsep
        \itshape
    #1\@addpunct{.}]\ignorespaces
}{%
  \popQED\endtrivlist\@endpefalse
  \addvspace{3pt plus 3pt} 
}

\crefname{section}{Sec.}{Sec.}
\crefname{theorem}{Thm.}{Thm.}
\crefname{lemma}{Lem.}{Lem.}
\crefname{corollary}{Col.}{Col.}
\crefname{table}{Tab.}{Tab.}
\crefname{algorithm}{Alg.}{Alg.}

\renewcommand\footnotetextcopyrightpermission[1]{}

\makeatletter
\patchcmd{\@algocf@start}
  {-1.5em}
  {0pt}
{}{}
\makeatother

\begin{document}

\fancyhead{}

\title{Parallel Strong Connectivity Based on Faster Reachability}



\author{Letong Wang}
\affiliation{%
  \institution{UC Riverside}
  \country{}
}
\email{lwang323@ucr.edu}

\author{Xiaojun Dong}
\affiliation{%
  \institution{UC Riverside}
  \country{}
}
\email{xdong038@ucr.edu}

\author{Yan Gu}
\affiliation{%
  \institution{UC Riverside}
  \country{}
}
\email{ygu@cs.ucr.edu}

\author{Yihan Sun}
\affiliation{%
  \institution{UC Riverside}
  \country{}
}
\email{yihans@cs.ucr.edu}


\begin{abstract}
Computing strongly connected components (SCC) is among the most fundamental problems in graph processing.
As today's real-world graphs are getting larger and larger, parallel SCC is increasingly important.
SCC is challenging in the parallel setting and is particularly hard on large-diameter graphs.
Many existing parallel SCC implementations can be even slower than Tarjan's sequential algorithm on large-diameter graphs.

To tackle this challenge, we propose an efficient parallel SCC implementation using a new parallel reachability algorithm.
Our solution is based on a novel idea referred to as vertical granularity control (VGC).
It breaks the synchronization barriers to increase parallelism and hide scheduling overhead.
To use VGC in our SCC algorithm, we also design an efficient data structure called the \emph{parallel hash bag}.
It uses parallel dynamic resizing to avoid redundant work in maintaining frontiers (vertices processed in a round).

We implement the parallel SCC algorithm by Blelloch et al.\ (J.\ ACM, 2020) using our new parallel reachability algorithm.
We compare our implementation to the state-of-the-art systems, including \gbbs{}, \ispan{}, \multistep{}, and our highly optimized Tarjan's (sequential) algorithm, on 18 graphs, including social, web, \knn{}, and lattice graphs.
On a machine with 96 cores, our implementation is the fastest on 16 out of 18 graphs.
On average (geometric means) over all graphs, our SCC is 6.0$\times$ faster than the best previous parallel code (\gbbs), 12.8$\times$ faster than Tarjan's sequential algorithms, and 2.7$\times$ faster than the \emph{best existing implementation on each graph}.

We believe that our techniques are of independent interest. We also apply our parallel hash bag and VGC scheme to other graph problems, including connectivity and least-element lists (LE-lists). Our implementations improve the performance of the state-of-the-art parallel implementations for these two problems.
\end{abstract}



\maketitle

\section{Introduction}
\label{sec:intro}
Computing strongly connected components (SCCs) is among the most fundamental problems in graph processing.
Given a directed graph $G=(V,E)$, we denote $n=|V|$ and $m=|E|$. We use $D$ as the diameter of $G$.
For simplicity, we assume $m\ge n$, but all algorithms in this paper works for any $n$ and $m$.
For two vertices $v,u\in V$, we use $u\leadsto v$ to denote that a path exists from $u$ to $v$.
Two vertices $v$ and $u$, are \defn{strongly connected} if $u\leadsto v$ and $v\leadsto u$.
An SCC is a maximal set of vertices on the graph that are strongly connected.
The SCC problem is to compute a mapping from each vertex to a unique label for its strongly connected component.
Computing SCC is useful in many applications inside and outside computer science,
and is widely used in database and data management applications~\cite{trissl2007fast,agrawal1990hybrid,randall2002link}.
For example, analyzing SCC on social networks can identify communities~\cite{mislove2007measurement} or estimate influence propagation~\cite{ohsaka2014fast}.
Applying SCC on the \knn{} graph (each point links to its $k$-nearest neighbors) for spatial data points is widely used in unsupervised learning~\cite{sharma2017knn,shekhar2018high,li2022robust}.
Many other fundamental problems in CS use SCC as an important primitive, such as graph matching~\cite{fan2010graph}, topological sort~\cite{cheng2013tf,sacharidis2009topologically}, graph contraction~\cite{clrs}, and code analysis~\cite{rugina2000symbolic}.
SCC is also widely-used on various types of graphs in other disciplines, e.g.,
lattice graphs~\cite{de2018percolation} in material science, on E.\ coli network~\cite{gama2016regulondb} in computer biology~\cite{morone2020fibration},
on food web analysis in ecology systems~\cite{allesina2005ecological}, and more~\cite{nadini2021mapping}.

Sequentially, Kosaraju's algorithm~\cite{aho1983data} and Tarjan's algorithm~\cite{tarjan1972depth} find all SCCs in a graph in $O(m)$ work (number of operations).
However, sequential algorithms can be slow to process today's large real-world graphs.
For example, on the Hyperlink12~\cite{webgraph} graph with billions of edges, Tarjan's SCC algorithm finishes in more than half an hour (see \cref{tab:graphfull}).
Hence, it is of great importance to seek \emph{parallel SCC solutions}.
Although SCC is also widely studied in parallel (see the literature review in~\cref{sec:related}), most existing algorithms are optimized based on two assumptions for the input graph: (1) with a low diameter and/or (2) with one large SCC.
Many algorithms (e.g., \ispan{}~\cite{ji2018ispan} and \multistep{}~\cite{slota2014bfs}) first use breadth-first searches (BFS) to identify the largest SCC and then run a subsequent \emph{coloring} phase to find all other SCCs with $O(m'D)$ work.
Here $m'$ is the edges not in the largest SCC.
When either of the assumptions is not satisfied, these algorithms will incur significant extra work than the $O(m)$ sequential algorithms, which cannot be compensated by parallelism.
Another solution in the \gbbs{} library~\cite{gbbs2021,dhulipala2020graph} implements the BGSS algorithm~\cite{blelloch2016parallelism}.
BGSS uses $\log n$ rounds of multi-reachability searches, and each of them starts from a subset of source vertices $S$ and finds all pairs of vertices $(v,s)$, where $s\in S$ and $s\leadsto v$.
BGSS has $O(m\log n)$ work that is independent of the two assumptions.
However, the \gbbs{} implementation uses parallel BFS to implement reachability searches, which process vertices with the same distance (referred to as a \defn{frontier}) in each round.
This approach incurs $O(D)$ rounds of global synchronization.
As each of the synchronization is costly, this implementation incurs high scheduling overhead and can be slow when $D$ is large.

Unfortunately, many applications of SCC have sparse input graphs with a large diameter $D$. For example, the \knn{} graphs used in unsupervised learning~\cite{sharma2017knn,shekhar2018high,li2022robust} and lattice graphs used in computational chemistry~\cite{de2018percolation} can have $D=\Theta(\sqrt{n})$ (see \cref{tab:graphfull}).
We tested existing SCC algorithms on graphs with both small diameters (e.g., social networks) and large diameters (e.g., \knn{} and lattice graphs), all widely used in real-world SCC applications.
\cref{fig:heatmap} shows that existing SCC algorithms work well for certain social networks but perform badly on other graphs.
On average, for \knn{} and lattice graphs, all existing parallel SCC algorithms on a 96-core machine are slower than the sequential Tarjan's algorithm.

\begin{figure*}
  \centering
  \includegraphics[width=2.1\columnwidth]{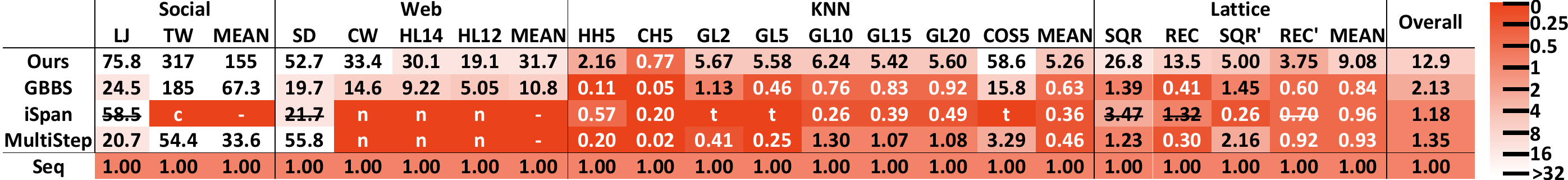}
  \caption{\small \textbf{The heatmap of relative speedup for parallel SCC algorithms over the sequential algorithm using 96 cores (192 hyperthreads).}
  \revision{Larger/white background means better.}
  ``Seq''$=$ Tarjan's algorithm~\cite{tarjan1972depth}. The numbers indicate how many times a parallel algorithm is faster than Tarjan's sequential algorithm ($<1$ means slower).  The three baseline algorithms~\cite{gbbs2021,ji2018ispan,slota2014bfs} are introduced in \cref{sec:exp}.
  ``t''$=$ timeout ($>5$ hours). ``c''$=$ crash. ``n''$=$ no support.
  Strikethrough numbers mean a wrong number of SCCs reported. 
  Complete results are in \cref{tab:graphfull}. \label{fig:heatmap}}
  \vspace{-2.5em}
\end{figure*}

In this paper, we propose \defn{an efficient SCC implementation with high parallelism on a wide range of graphs}. 
We also use the BGSS algorithm to bound the work.
The core of our idea is to improve parallelism by avoiding $O(D)$ rounds of synchronization in reachability searches and thus reducing the scheduling overhead.
To do this, we propose a novel idea referred to as the \defn{vertical granularity control} (VGC) optimization.
The high-level idea of VGC is to break the synchronization barriers and increase parallelism.

Particularly, for the reachability queries, unlike parallel BFS that only visits the neighbors of the vertices in the frontier, we want to visit a much larger set of vertices that can be multiple hops away.
This is achieved by a ``local search'' algorithm that allows each vertex in the frontier to visit more than direct neighbors in one round.
We will discuss more details of VGC and local search in \cref{sec:local,sec:tau}.
This approach saves most synchronization rounds and improves the performance, especially on large-diameter graphs.

We note that one technical difficulty in \VGC{} and local search is to handle the non-determinism in generating the next frontier---all vertices in the frontier explore their proximity in parallel, in a random order decided by the runtime scheduling.
This disables the ``edge-revisit'' approach in existing BFS algorithms~\cite{gbbs2021}, which first process the frontier to count the size of the next frontier and preallocate memory and then revisit the frontier to output vertices to the next frontier. We present more details about this challenge in \cref{sec:prelim,sec:hashbag}.
To maintain the frontier more efficiently (and correctly) in \VGC{}, we 
propose a novel data structure called the \defn{parallel \hashbag{}}.
It supports efficient insertion and extract-all operations for an unordered set structure and dynamic resizing in an efficient manner.
Our parallel \hashbag{} is similar to a resizable hash table but avoids copying on resizing.
We maintain a multiple-level structure in exponentially growing sizes.
We use atomic operations to enable concurrent insertions and a sampling scheme to support resizing.
Our \hashbag{} is theoretically-efficient and fast in practice.
We describe the details of the \hashbag{} in \cref{sec:hashbag}. 

Using these techniques, our new SCC approach achieves good performance on various graphs (see \cref{fig:heatmap}), all widely used in real-world, including
the \emph{largest publicly available graph (HL12)} with 3.6 billion vertices and 128 billion edges.
Among all implementations, our SCC achieves the best performance on 16 out of 18 graphs.
As expected, our SCC performs particularly well on large-diameter graphs (\knn{} and lattice graphs).
Compared to \emph{the fastest running time among existing parallel algorithms on each graph},
our implementation is 2.3--12$\times$ faster on large-diameter graphs, and up to 3.7$\times$ faster on low-diameter graphs.
Our results show that the good performance of our SCC mainly comes from \VGC{} (using 3--200$\times$ fewer rounds than BFS, see \cref{fig:SCC-rounds}) and \hashbag{}s (see \cref{fig:SCC_Total_Break}).  We release our code on github~\cite{scccode}.

We believe that our proposed techniques are general and can be applied to many graph algorithms.
As proofs-of-concept, we apply the proposed techniques to two more algorithms: connected components (CC) and least-element lists (LE-lists).
Using our new techniques, our solutions outperform the state-of-the-art algorithms by up to 3.2$\times$ on CC and 10$\times$ on LE-lists.
We overview these problems in \cref{sec:others} and present the experimental results in \cref{sec:exp-cc-lelists}.

We summarize our contributions as follows.

\begin{enumerate}[wide,topsep=0pt]
  \item Two general techniques (\vgc{} and parallel \hashbag{}) to optimize the performance of graph traversal.
  \item Fast implementations on SCC, CC, and LE-lists using the proposed techniques.
  Our SCC algorithm greatly outperforms existing implementations.
  \item In-depth experimental evaluation of the problems and the optimizations in this paper across multiple types of graphs.
\end{enumerate}

\hide{
work mostly focuses on optimizing the performance on low-diameter graphs (e.g., social networks).
Unfortunately, in many applications of SCC, the input graph has large diameter.
For example, \knn{} graphs are widely used in clustering, and they usually have diameters around $\sqrt{n}$ (see \cref{tab:graphfull}).
On real-world datasets,
their diameters can be more than thousands in contrast to (typically) 10-100 on social networks.
In our experiments, we observe that existing parallel SCC implementations (e.g.,~\cite{hong2013fast,ji2018ispan,slota2014bfs,gbbs2021}) \emph{suffer from low parallelism on large-diameter graphs}.
\cref{fig:heatmap} shows the speedup of \emph{parallel SCC} implementations over the \emph{sequential algorithm} on four types of graphs: social networks, web graphs, \knn{} graphs,
and lattice graphs (see details of the graphs in \cref{tab:graphfull}), tested on a machine with 96 cores.
All social, web and \knn{} graphs are from \emph{real-world} datasets, and lattice graphs are randomly generated.
On the low-diameter graphs (social and web graphs), almost all implementations achieve reasonable speedup over the sequential algorithm.
However, on large-diameter graphs (\knn{} and lattice graphs), \defn{none of the existing parallel implementations has a better overall performance than Tarjan's sequential algorithm}.
We review more details of related work in \cref{sec:related}.

This paper studies efficient parallel implementation for strongly connected components (SCC),
especially addressing the challenge of achieving good parallelism on large-diameter graphs.
The performance bottleneck of parallel SCC on large-diameter graphs
is due to the long dependency chain to propagate the ``reachability'' information
from some sources along the paths.
There are two major approaches in the literature.
The \emph{coloring} approach, used in \ispan{}~\cite{ji2018ispan} and \multistep{}~\cite{slota2014bfs},
is to let the vertices propagate their labels to their reachable neighborhoods.
This step is used after finding the largest SCC (referred to as $\sccone$), and incurs extra $\Omega(D^{*}n)$ work,
where $D^{*}$ is the diameter of the subgraph without $\sccone$.
When the graph has a large $\sccone$, and the rest of vertices are not highly connected, coloring can have reasonable good performance.
However, on many large-diameter graphs (e.g., the \knn{} graphs shown in ~\cref{tab:graphfull}) where $\sccone$ is small,
these algorithms need to tackle a large portion of the graph in the coloring step, and thus can be inefficient.
The second type of approach is used in BGSS algorithm~\cite{} (implemented in the \gbbs{} library),
which compute all reachable vertices from a set of source(s) using breadth-first search (BFS)
to bound the work in $O(m\log n)$ in expectation.
However, BFS traverses all vertices in rounds based on hop distances from the source(s).
and incurs high scheduling overhead to synchronize threads between rounds.
This severely limits the parallelism of this implementation.

We also use the BGSS algorithm to bound the work, and use novel approaches to achieve high parallelism
on \emph{all types of graphs}. Our main observation is that, in SCC, we do not need the \emph{BFS ordering} (distance information) but only the \emph{reachability} information on whether a search can reach a vertex or not.
This relaxation enables effective optimizations.
We propose two techniques:
1) the \defn{local search}, which is an optimization for reachability search to reduce the costs of synchronizing threads,
and 2) the \defn{\hashbag{}}, which is a data structure to efficiently maintain the vertices processed in each round (called the \defn{frontier} of each round).

Our first technique is the \defn{local search} optimization in reachability queries.
When processing a vertex $v$, the BFS algorithm visits all neighbors of $v$, pushing out the frontier by one hop from the source.
On large-diameter graphs, this incurs a lot of rounds to finish,
and the scheduling overhead (e.g., forking and synchronizing threads between rounds) is much more costly than the
actual computation, as each thread usually only visits a small number of vertices.
Our local search allows for more asynchrony in processing the frontiers,
which effectively reduce the number of rounds and saves the synchronization cost.
When processing a vertex $v$, in addition to the direct neighbors of $v$, we maintain a fixed-size buffer for local search in the local (stack) memory and further search outwards until the buffer is full.
The vertices may not be visited in the BFS order, 
but this does not affect the reachability of the vertices.
Although this idea sounds simple, we note that this \emph{also requires to significantly change the implementation to maintain the frontiers},
which we tackle by designing the new data structure of \emph{hash bag} (see more details below).
Local search significantly improves the performance of our algorithms (e.g., 2.3--14$\times$ on \knn{} and lattice graphs).
We discuss more details in \cref{sec:local}.

Our second technique is a data structure, called the \defn{parallel \hashbag{}}, to maintain the vertex set being processed (the \emph{frontier}) in each round.  
The goal is to avoid the redundant work in processing the frontier multiple times in order to count the size of the next frontier to allocate memory.
This cost can be more significant (and much more complicated) when local search is enabled.
This is because local search can explore several hops away from a vertex $v$ in the current frontier, and
we no longer have a fixed set of ``candidates'' for the next frontier
(originally, the candidates are just all neighbors of $v$).
Therefore, we need a data structure that resizes dynamically and supports efficient insertion and list-all operations.
Our parallel \hashbag{} is similar to a resizable hash table but avoids copying on resizing.
We maintain a multiple-level structure in exponentially growing sizes without sacrificing the cost per operation.
We use atomic operations to enable concurrent insertions and a sampling scheme to support resizing,
such that we do not need to know the exact size of the next frontier.
Our \hashbag{} is theoretically-efficient 
and fast in practice.
We describe the \hashbag{} in detail in \cref{sec:hashbag}. 

We believe that our proposed techniques are general and of independent interest. In addition to the SCC problem, which is our main focus, we apply the proposed techniques to two more algorithms: connected components (CC) and least-element lists (LE-lists). Using our techniques, we improve the performance of a state-of-the-art CC algorithm and present the first parallel LE-lists implementation.
We overview these problems in \cref{sec:others} and present some experimental results in \cref{sec:exp-cc-lelists}.
}





\hide{
However, BFS itself is hard to parallelize for arbitrary graphs, which 
results in theoretically-inefficient implementations for SCC---the span of BFS is proportional to the diameter of the graph.
For instance, as shown in \cref{fig:heatmap}, in GBBS~\cite{gbbs2021},
the BFS-based SCC algorithms achieve very good performance on low-diameter graphs (e.g., social and web graphs)
but do not perform well on larger-diameter graphs (e.g., \knn{} and lattice),
and in many cases are even slower than the standard \emph{sequential} algorithms.
The reason is that on large-diameter graphs, BFS requires more rounds and visits fewer vertices in each round, which causes insufficient parallelism in each round and significant overhead for global synchronization between rounds.
Taking the geometric means on all graphs,
\gbbs{}'s SCC on 96 cores is only 2x faster than the standard \emph{sequential algorithm} (see \cref{fig:GL5COS5_rounds}).}



\hide{
As a result, there were no existing SCC implementations with desirable performance on all graph instances tested in this paper.

Our main observation is that, in SCC and CC problems, we do not need the \emph{BFS ordering} (distance information) but only need the \emph{reachability} information on whether a search can reach a vertex or not (formal definition given in \cref{sec:technique}).
This relaxation enables effective optimizations in parallel implementations.
The correlation between reachability and connectivity-related problems has been raised in recent theory work~\cite{blelloch2016parallelism,schudy2008finding,fineman2019nearly}.
In this paper, we further study two questions to make the idea highly practical.
First, given that we do not need to maintain the BFS ordering, how can we design a parallel reachability algorithm?
Second, how can we use parallel reachability to design connectivity-related algorithms?
}
\hide{
The main contribution of this paper is an efficient reachability algorithm with highly-practical optimizations.
Note that this includes both single-reachability from one source and a more complicated case of multi-reachability from a set of sources,
which is used in the parallel SCC algorithm~\cite{blelloch2016parallelism}.
Our reachability algorithm is based on the parallel round-based BFS algorithm, and we show two algorithmic improvements.

Our first technique is an efficient \emph{local search} optimization in reachability queries.
When processing a vertex $v$, the BFS algorithm visits all neighbors of $v$, pushing out the frontier by one hop from the source.
On large-diameter graphs, this incurs a lot of rounds to finish, between which global synchronization is needed.
In this case, the scheduling overhead (e.g., forking and synchronizing threads) is much more costly than the
actual computation, as each thread usually only visits a small number of vertices.
Our local search aims at reducing the number of rounds and thus saving the synchronization cost.
When processing a vertex $v$, in addition to the direct neighbors of $v$, we maintain a fixed-size buffer for local search in the local (stack) memory and further search outwards until the buffer is full.
This may not visit the vertices in the BFS order, 
but this does not affect the reachability of the vertices.
Conceptually, we believe this optimization functions as a parallel granularity control, such that we try to \emph{let each thread perform a reasonably large amount of work},
and is also similar to adding shortcuts to the graph on the fly to reduce the number of rounds needed.
Surprisingly, we were unaware of this simple yet effective optimization previously,
which significantly improves the performance of our algorithms (e.g., 2-7x for SCC on \knn{} graphs).
We discuss more details in \cref{sec:local}.

Second, we propose a data structure, referred to as the \emph{parallel \hashbag{}}, to maintain the vertex set being processed (the \emph{frontier}) in each round efficiently.
The goal is to avoid the overhead in counting the size of the next frontier to allocate memory, which can result in redundant work in processing the frontier multiple times.
Our parallel \hashbag{} is similar to a resizable hash table but avoids copying on resizing.
We maintain a multiple-level structure in exponentially growing sizes without sacrificing the cost per operation.
Our design uses atomic operations to enable concurrent insertions and a carefully-designed sampling scheme to support resizing at the same time,
such that we do not need to know the exact size of the next frontier.
Our \hashbag{} is theoretically efficient regarding work (total operations) and span (longest dependent operations) and fast in practice.
Using parallel \hashbag{}s to maintain the frontiers in SCC significantly improves the performance of these algorithms.
We describe the \hashbag{} in detail in \cref{sec:hashbag}. \hashbag{}s can be applied to most edge-map based algorithms, such as CC and LE-list, which simplifies coding and improves the performance.

To answer the second question about applying reachability algorithm to the connectivity-related problems,
we have to carefully choose or redesign the relevant algorithms to decouple the features enabled by BFS
but still guarantee the correctness and cost bound of the algorithms.
All the algorithms we propose or implement in this paper have a theoretical guarantee and are highly practical.
For SCC, we plug our reachability algorithm into the BGSS algorithm~\cite{blelloch2016parallelism}
, which is theoretically efficient (wrt.\ the number of dependent reachability searches~\cite{blelloch2020randomized}).
For connectivity, we also apply our techniques to an efficient algorithm from ConnectIt~\cite{dhulipala2020connectit} and further improve the performance.
Meanwhile, we show this algorithm is theoretically efficient (the only known theoretically efficient CC algorithm~\cite{SDB14} is not as fast in practice).
We also consider the least-element lists (LE-lists)~\cite{blelloch2016parallelism}, which is a widely-used graph building block in many applications (e.g.,~\cite{cohen2004,du2013,blelloch2017efficient,Khan2008}).
We show the first parallel implementation for LE-lists based on the BGSS LE-lists algorithm~\cite{blelloch2016parallelism} using the proposed techniques.
}


\hide{
\begin{figure}
  \centering
  \includegraphics[width=\columnwidth]{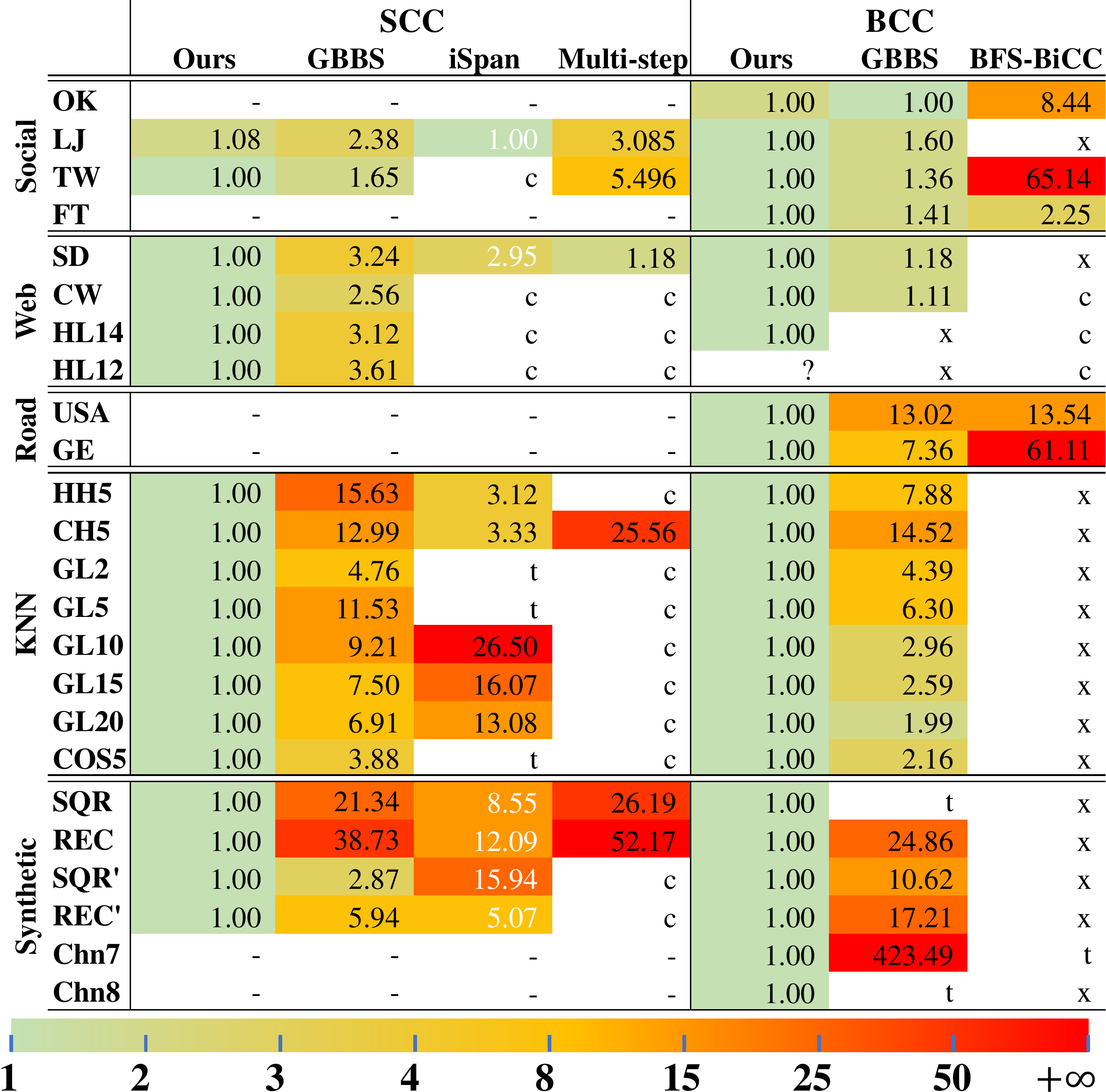}
  \caption{Heatmap}\label{fig:heatmap}
\end{figure}
}

\hide{

Figure \cref{fig:intro1} shows the performance of GBBS, which is one of the best existing implementations for SCC and BCC.
On large social-networks, both algorithms perform reasonably well.
However on the same size of graphs with large diameter, such as road network and KNN graphs, they requires much longer running time, and for some of them even longer than processing a much larger social network.
The problem lies in that existing algorithms for BC or SCC rely on Breadth First Search (BFS) as a core component, which itself is hard for large-diameter graphs.
State-of-the-art parallel BFS algorithm processes vertices in rounds. The $i$-th round processes all $i$-hop neighbors of the source vertex in parallel, called the current \emph{frontier}, explores their neighbor sets and puts the newly visited neighbors in the next frontier.
However, BFS itself is hard to parallelize for large diameter graphs.
There are two major issues.
First of all, to generate the next frontier in a consecutive array, the algorithm needs to visit the neighborhood of the current frontier at least twice. The first time selects the appropriate vertices (excluding the visited ones or duplicates) and compute the offset for each selected vertices, and the second time writes (packs) the selected vertices to the corresponding slots in the next frontier. This creates extra overhead in the algorithm.
For dense graphs, BFS algorithms can use a \emph{dense mode} (see details in xx) to avoid this second-round packing, but this only pays off when the frontier is large (called the dense rounds).
Secondly, on sparse graphs, the frontier size is usually small, and the number of total rounds needed may be large.
Since parallelism is explored only within each round, synchronizations are needed between rounds.
Therefore, large number of rounds directly leads to longer running time.
Again this is not an issue for dense graphs, since their diameters are small, the number of rounds is usually a small number.

Our design is based on an observation that such data structures for maintaining frontiers usually requires fast operations of
\fname{put} and \fname{forall} (in parallel), whereas a fast look up operation is not required.
Our \hashbag{} is implemented by a pre-allocated array with sufficiently large size (usually just the number of vertices in the graph),
but only uses a small prefix of size $O(s)$, where $s$ is the current number of elements in the bag.
An element is put into the bag by scattering it to an arbitrary slot in the current array, and resizing is done by swinging the tail pointer to twice of the
current size. To decide when to resize, we carefully design a sampling scheme to estimate the current size of the bag.
By putting all vertices in the next frontier to the \hashbag{}, we can reconstruct the frontier by visiting the prefix of the bag and skipping the empty slots.
This visits at most $O(s)$ elements for a frontier of size $s$, instead of visiting the full edge list of the previous frontier.

}

\hide{
There are two major difficulties
First of all, to generate the next frontier in a consecutive array, the algorithm needs to visit the neighborhood of the current frontier at least twice. The first time selects the appropriate vertices (excluding the visited ones or duplicates) and compute the offset for each selected vertices, and the second time writes (packs) the selected vertices to the corresponding slots in the next frontier. This creates extra overhead in the algorithm.
For dense graphs, BFS algorithms can use a \emph{dense mode} (see details in xx) to avoid this second-round packing, but this only pays off when the frontier is large (called the dense rounds).
Secondly, on sparse graphs, the frontier size is usually small, and the number of total rounds needed may be large.
Since parallelism is explored only within each round, synchronizations are needed between rounds.
Therefore, large number of rounds directly leads to longer running time.
Again this is not an issue for dense graphs, since their diameters are small, the number of rounds is usually a small number.
}

\hide{
Given the large size of today's real-world graphs, it is of great importance to seek \emph{parallel solutions} for these problems.
Although there exist
classic sequential algorithms to solve SCC efficiently, such as Kosaraju's algorithm~\cite{aho1983data} and Tarjan's algorithm~\cite{tarjan1972depth} (both in $O(m)$ cost),
they both use depth-first search, which is not parallelizable~\cite{reif1985depth}.
On large real-world graphs, the sequential algorithms can be slow
(e.g., Tarjan's algorithm takes half an hour to process the Hyperlink 12 graph~\cite{}). }


\section{Preliminaries}\label{sec:prelim}

\hide{In this section, we first define the model and notation we use to analyze algorithms.
Then we review the concept of \emph{reachability search} that is widely used in parallel SCC algorithms.
Finally we review the BGSS algorithm~\cite{blelloch2016parallelism},
and the implementation in the GBBS library~\cite{dhulipala2021theoretically}.
Our algorithm is based on the BGSS algorithm,
but with novel optimizations to improve the performance.
}

\begin{figure*}
  \centering
  \vspace{-1em}
  \includegraphics[width=2.1\columnwidth]{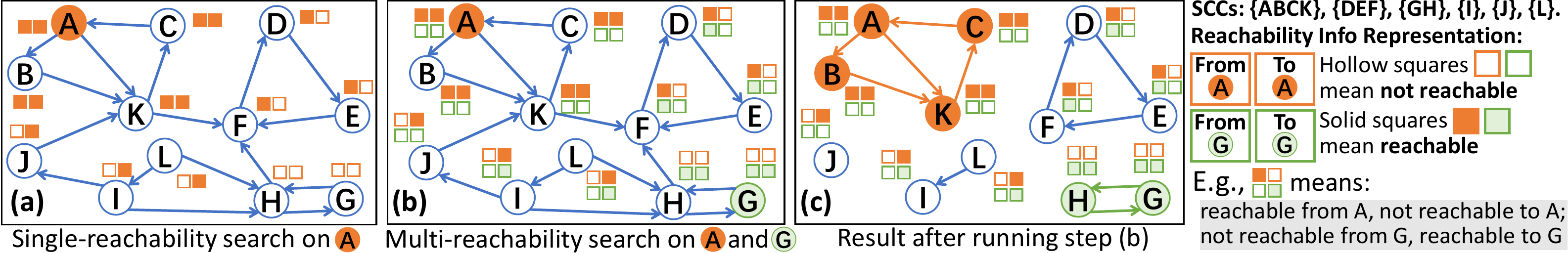}
  \caption{\small \textbf{Example of single- and multi-reachability searches and reachability-based SCC algorithms. }
  The SCCs are $\{A,B,C,K\}$, $\{D,E,F\}$, $\{G,H\}$, $\{I\}$, $\{J\}$, $\{L\}$.
  (a). Single-reachability search results on $A$. (b). Multi-reachability search results on $A$ and $G$.
  The reachability information (referred to as the \emph{signature}) of a vertex is shown as a $2\times 2$ lattice.
  (c). After running multi-reachability searches from $A$ and $G$, we can identify two SCCs: all orange vertices (reachable from and to $A$) and all green vertices (reachable from and to $G$). 
  Two vertices are in the same SCC only if they have the same signature. Thus, all cross edges (endpoints with different signatures) are removed. Later steps in the SCC algorithm only need to process the blue vertices and edges in (c).
  }\label{fig:reachability}
  \vspace{-.3in}
\end{figure*} 

\myparagraph{Notations and Computational Model.}
We use $\tilde{O}(f(n))$ to denote $O(f(n)\cdot \mathit{polylog}(n))$.
We use $O(f(n))$ \defn{with high probability} (\whp{}) in $n$ to demote $O(kf(n))$ with probability at least $1-n^{-k}$ for $k \geq 1$. We omit ``in $n$'' with clear context.
For a graph $G=(V,E)$, we denote $n=|V|$, $m=|E|$, and $D$ as the diameter of the graph.

This paper uses
work-span (or work-depth) model for nested parallelism based on fork-join and binary forking~\cite{clrs,blelloch2020optimal} for theoretical analysis.
We assume a set of \thread{}s that share the memory.
Each \thread{} acts like a sequential RAM, plus a \forkins{} instruction to create two new child threads running in parallel.
When both children finish, the parent thread continues, implying a synchronization.
The computation can be executed by a randomized work-stealing scheduler in practice.
We assume unit-cost atomic operation \cas{}$(p,v_{\mathit{old}},v_{\mathit{new}})$ \revision{(or \CAS{})},
which atomically reads the memory location pointed to by $p$, and write value $v_{\mathit{new}}$ to it if the current value is $v_{\mathit{old}}$.
It returns $\true{}$ if successful and $\false{}$ otherwise.
An algorithm's \defn{work} is the total number of instructions and
the \defn{span} (depth) is the length of the longest sequence of dependent instructions in the computation.
Our algorithm uses the scheduler in Parlaylib~\cite{blelloch2020parlaylib} to support fork-join parallelism.





\myparagraph{Reachability Search.} Given a directed graph $G=(V,E)$ and a set of source vertices $S\subseteq V$, a (forward) reachability search finds the set of reachable pairs from any source $\{(s,t): s\in S, t\in V, s \leadsto t\}$.
We say a vertex $v$ is \defn{reachable from} $s$ if $s \leadsto v$, and \defn{reachable to} $s$ if $v\leadsto s$.
We define \emph{backward reachability search} as searching in the reverse direction of the edges (i.e., on the transpose graph $G^T$). 
We use \defn{single-reachability search} to refer to the case where $|S|=1$, and use \defn{multi-reachability search} otherwise.
\cref{fig:reachability} illustrates single- and multi-reachability searches.

Many SCC algorithms are based on reachability searches~\cite{coppersmith2003divide,blelloch2016parallelism,schudy2008finding,barnat2011computing,ji2018ispan,slota2014bfs,tomkins2014sccmulti}. 
The key idea is that 1) all vertices reachable both from and to a vertex $v$ are in the same SCC, and
2) two vertices are in the same SCC iff.\ they have the same reachability information from \emph{any source}.
Therefore, an edge $(u,v)$ is not in any SCC (and can be removed) if $u$ and $v$ have different reachability information from \emph{any source}.
We call such edges \defn{cross edges}.
\hide{
Consider the multi-reachability search in \cref{fig:reachability}(b) on $A$ and $G$.
$\{A,B,C,K\}$ are reachable both from and to $A$;
$\{H,G\}$ are reachable both from and to $G$. Therefore, $\{A,B,C,K\}$ and $\{G,H\}$ are identified as two SCCs (the orange and green components in \cref{fig:reachability}(c)).
In \cref{fig:reachability}, we use a $2\times 2$ lattice to represent the reachability information of a vertex, and call it the \defn{signature} of it.
An edge connecting two vertices with different signatures is a cross edge, and can be removed in \cref{fig:reachability}(c).
We only need to process the blue vertices and edges in \cref{fig:reachability}(c) in later parts of the algorithm.
}
We present an example in \cref{fig:reachability}.
We use a $2\times 2$ lattice as the reachability information of a vertex (called the \defn{signature} of it).
An edge connecting two vertices with different signatures is a cross edge, and can be removed.


Some theoretical results show that a single-reachability search can be done in $\tilde{O}(m)$ work and $o(n)$ span~\cite{fineman2019nearly,jambulapati2019parallel}.
However, the hidden terms in the bounds are large, and these algorithms are unlikely to be practical.
In practice, BFS with $O(m)$ work is used for reachability searches~\cite{slota2014bfs,gbbs2021}.
It works well on low-diameter graphs but performs poorly on large-diameter graphs.
We review parallel BFS algorithms for reachability search later in this section.

\begin{algorithm}[t]
\small
\fontsize{8pt}{8.7pt}\selectfont
\caption{The BGSS algorithm for parallel SCC~\cite{blelloch2016parallelism}\label{alg:SCC}}
\KwIn{A directed graph $G=(V,E)$} 
\KwOut{The component label $L[\cdot]$ of each vertex.}
\SetKwFor{parForEach}{parallel\_for\_each}{do}{endfor}
\SetKwInOut{Maintains}{Maintains}
\DontPrintSemicolon
	$L \gets \{-1, \dots, -1\}$ 	\label{line:initialize} \\
    Partition $V$ into $\log n$ batches $P_{1..\log n}$, where $|P_{i}|=2^{i-1}$\label{line:scc_partition}\\
    $V'\gets V$\\
	\For{$i \gets 1, \dots , \log n$} { 									\label{lst:line:for}
		$\frontier{} \gets \{v \in P_i \cap V'\}$ 	\Comment{Initial frontier}\label{lst:line:Frontier}\\	
        \tcp{\mf{MultiReach} skips an edge $(u,v)$ if $L(u)\ne L(v)$}
		$L_{\mathit{out}} \gets \mf{MultiReach}(G, L, \frontier{})$ \Comment{Forward reachable pairs}	\label{line:scc:reach1}\\
		$L_{\mathit{in}} \gets \mf{MultiReach}(G^T, L, \frontier{})$ \Comment{Backward reachable pairs}\label{line:scc:reach2}\\
        \parForEach {$i \in V'$\label{line:scc:reach3}} {
            $R_1\gets \{v~|~(v,i) \in L_{\mathit{in}}\}$\Comment{Reachable to $i$}\\
            $R_2\gets \{v~|~(v,i) \in L_{\mathit{out}}\}$\Comment{Reachable from $i$}\\
            \lIf {$R_1\cap R_2\ne\emptyset$} {
                $L[i]\gets\max_{v\in R_1\cap R_2}v$\label{line:scc:label1}
            }
            \lElse {
                $L[i]\gets \mathit{hash}(L[i],R_1,R_2)$\label{line:scc:label2}
            }
        }

        $V'\gets V'\setminus \{i~|~\exists (v,i)\in L_{\mathit{in}} \cap L_{\mathit{out}}\}$\label{line:scc:reach4}\\
	}
	\Return{$L$}
\end{algorithm}

\hide{
\begin{algorithm}[t]
\small
\caption{The BGSS algorithm for parallel SCC~\cite{blelloch2016parallelism}\label{alg:SCC}}
\KwIn{A directed graph $G=(V,E)$} 
\KwOut{The component label $L[\cdot]$ of each vertex.}
\SetKwFor{parForEach}{parallel\_for\_each}{do}{endfor}
\SetKwInOut{Maintains}{Maintains}
\DontPrintSemicolon
	$L \gets \{0, \dots, 0\}$ 	\label{line:initialize} \\
	$V' \gets \{v \in V~|~ \mathit{deg}_{+} > 0 \land \mathit{deg}_{-} > 0\}$ \Comment{Trimming}\label{line:trim}\\
    Random pick a source $s_1$\\
	$V_{\mathit{forward}} \gets \mf{SingleReach}(s_1,G)$ \label{line:firstscc1}\\			
	$V_{\mathit{backward}} \gets \mf{SingleReach}(s_1,G^T)$ \Comment{$G^T$ reverses edges in $G$}\label{line:firstscc2}\\
    $\mathit{SCC}_1\gets V_{\mathit{forward}} \cap V_{\mathit{backward}}$ \Comment{First SCC}\\
	\lparForEach {$v\in \mathit{SCC_1}$} {$L[v]\gets s_1$\Comment{Label $\mathit{SCC}_1$}}
	$V' \gets V'\setminus\mathit{SCC}_1$\\
    Partition $V'$ into $\log n$ batches $P_{1..\log n}$, where $|P_{i}|=2^{i-1}$\label{line:scc_partition}\\
	\For{$i \gets 1, \dots , \log n$} { 									\label{lst:line:for}
		$\frontier{} \gets \{v \in P_i \cap V'\}$ 	\Comment{Initial frontier}\label{lst:line:Frontier}\\	
        \tcp{\mf{MultiReach} skips an edge $(u,v)$ if $L(u)\ne L(v)$}
		$L_{\mathit{out}} \gets \mf{MultiReach}(G, L, \frontier{})$ \Comment{Forward reachable pairs}	\label{line:scc:reach1}\\
		$L_{\mathit{in}} \gets \mf{MultiReach}(G^T, L, \frontier{})$ \Comment{Backward reachable pairs}\label{line:scc:reach2}\\
        \parForEach {$i \in V'$\label{line:scc:reach3}} {
            $R_1\gets \{v~|~(v,i) \in L_{\mathit{in}}\}$\Comment{Reachable to $i$}\\
            $R_2\gets \{v~|~(v,i) \in L_{\mathit{out}}\}$\Comment{Reachable from $i$}\\
            \lIf {$R_1\cap R_2\ne\emptyset$} {
                $L[i]\gets\max_{v\in R_1\cap R_2}v$\label{line:scc:label1}
            }
            \lElse {
                $L[i]\gets \mathit{hash}(R_1,R_2)$\label{line:scc:label2}
            }
        }

        $V'\gets V'\setminus \{i~|~\exists (v,i)\in L_{\mathit{in}} \cap L_{\mathit{out}}\}$\label{line:scc:reach4}\\
	}
	\Return{$L$}
\end{algorithm}
}




\myparagraph{The BGSS Algorithm.}
Our parallel SCC solution uses the BGSS SCC algorithm~\cite{blelloch2016parallelism} based on reachability searches, which is shown in \cref{alg:SCC}.
To achieve good parallelism while bounding the work, the BGSS algorithm uses $\log n$ batches of reachability searches.
The algorithm first randomly permutes the vertices and groups them into batches of sizes $1,2,4,8,...$ \revision{in a prefix-doubling manner
(the multiplier is not necessary to be $2$, but can be any constant $\beta>1$)}.
In the $i$-th round, the algorithm uses batch $i$ with $2^{i-1}$ vertices as the sources to run (forward and backward) multi-reachability searches, marks SCCs, and removes cross edges.
In this way, 
the BGSS algorithm takes $O(W_R(n,m)\log n)$ expected work and $O((D_R(n,m)+\log n)\log n)$ span \whp{},
where $W_R(n,m)$ and $D_R(n,m)$ are the work, and span for a reachability search on a graph with $n$ vertices and $m$ edges.
The BGSS algorithm was implemented by Dhulipala et al.\ as part of the \gbbs{} library~\cite{gbbs2021,dhulipala2020graph}, which uses (multi-)BFS for reachability searches (see more details later).
There are two SCC algorithms in \gbbs{}, and we refer to the \emph{RandomGreedy} version, as it is faster in most of our tests.

\myparagraph{Parallel BFS}.
\label{sec:bfs}
We briefly review parallel BFS, because it is used in previous work for reachability search, and some of the concepts are also used in our algorithm.
There are many BFS algorithms (e.g.,~\cite{beamer2015gap,zhang2018graphit,nguyen2013lightweight}).
We review the version in Ligra~\cite{shun2013ligra}, as it is widely-used in other graph libraries~\cite{dhulipala2017,dhulipala2020connectit,gbbs2021}, and more importantly, later extended to \defn{multi-BFS} that can be used in multi-reachabilty searches needed by our SCC algorithm.
We start with BFS from a \emph{single} source $s\in V$ (high-level idea in \cref{alg:BFS}).
The algorithm maintains a \defn{frontier} of vertices to explore in each round, starting from the source, and finishes in $D$ rounds.
In round $i$, the algorithm \emph{processes} (visits their neighbors) the current frontier $\ff_i$, 
and puts all their (unvisited) neighbors to the next frontier $\ff_{i+1}$.
If multiple vertices in $\ff_i$ attempt to add the same vertex to $\ff_{i+1}$, a \cas{} is used to guarantee that only one will \emph{succeed}.
In existing libraries~\cite{shun2013ligra,gbbs2021}, 
processing $\ff_i$ involves visiting all incident edges of $\ff_i$ twice.
The first visit decides the \emph{successfully} visited neighbors for each $v\in \ff_i$,
and assigns the right size of memory in $\ff_{i+1}$ for each of them.
The second visit lets each $v\in \ff_i$ write these neighbors to $\ff_{i+1}$.
We call this the \defn{edge-revisit} scheme, and we will show how our \emph{\hashbag{}} avoids the second visit and improve the performance.

This idea has been extended to \emph{multiple} sources $S\subseteq V$ with two changes~\cite{gbbs2021}.
First, a parallel hash table $T$~\cite{shun2014phase} is used to maintain the \defn{reachability pairs} $(v,s)$ where $v\in V$ and $s\in S$. 
Second, for each $v\in \ff_i$ and its successfully visited neighbor $u$, we find all pairs $(v,s)$ from the hash table $T$, and add $(u,s)$ to $T$ if $(u,s)\notin T$.
The number of reachability pairs generated by the BGSS algorithm is proved to be $O(n\log n)$ \whp{}~\cite{blelloch2021read}.


One challenge of using parallel BFS for reachability queries is
the large cost to create and synchronize threads between rounds,
which is especially expensive for large-diameter graphs (more rounds needed).
In this paper, we will show how our new techniques reduces
the scheduling overhead to achieve better parallelism.

\begin{algorithm}[t]
\small
\caption{Framework of Parallel BFS\label{alg:BFS}}
\KwIn{A directed graph $G=(V,E)$ and a source $s\in V$}
\SetKwFor{parForEach}{parallel\_for\_each}{do}{endfor}
\SetKwInOut{Maintains}{Maintains}
\DontPrintSemicolon
$\ff_0=\{s\}$\\
$i\gets 0$\\
\While {$\ff_{i}\ne \emptyset$}{
\hide{
\parForEach{ $v\in \ff_i$}{
  Put all unvisited neighbors of $v$ to $\ff_{i+1}$
}
}
Process all $v\in \ff_i$ and their edges in parallel, put all their unvisited neighbors (but avoid duplicates) to $\ff_{i+1}$\\
$i\gets i+1$
}
\end{algorithm}

\section{Fast Parallel Algorithm for Reachability}\label{sec:technique}

\hide{
As previously discussed, most previous parallel SCC algorithms (e.g., \ispan{}~\cite{ji2018ispan} and \multistep{}~\cite{slota2014bfs}) use graph coloring with $O(m'D)$ work, where $m'$ is the edges not in the largest SCC, and $D$ is the diameter of the graph.
If the input graph has a large diameter or does not have a large SCC, these algorithms perform badly as shown in \cref{fig:heatmap}.
The GBBS library~\cite{gbbs2021} implemented the BGSS algorithm~\cite{blelloch2016parallelism} that has $O(m\log n)$ work.
However, it uses parallel BFS to implement the reachability queries that incurs $O(D\log n)$ rounds of global synchronization.
As global synchronization is very costly, GBBS also suffers from poor performance on graphs with large diameters (\cref{fig:heatmap}).
}
To implement an efficient parallel SCC algorithm, we use the BGSS algorithm to bound the work,
and present novel ideas for fast reachability search to enable high parallelism.
In this section, we present two main techniques in this paper: the \defn{\vgc{} (\VGC{})} with the \defn{\hashbag{}} data structure.
Our \VGC{} optimization is designed to address the challenge of low parallelism in computing SCC on sparse and large-diameter graphs.
The goal is to enable a proper size for each parallel task to hide the scheduling overhead.
Our idea is to let each vertex search out multiple hops in each parallel task,
and thus the number of needed rounds in reachability searches is reduced.
The details of the local search is in \cref{sec:local}.
While the high-level idea sounds simple, this brings up the challenge of non-determinism---each vertex may explore multiple hops and the explored neighborhood depend on runtime scheduling, which results in some complication in generating the frontier by the edge-revisit scheme (see \cref{sec:bfs}).
Therefore, we propose a data structure called the \defn{parallel \hashbag{}}, to maintain the frontier more efficiently.
Our \hashbag{} is theoretically-efficient and fast in practice, and more details are given in \cref{sec:hashbag}.

Combining both techniques, we achieve fast single- and multi-reachability searches.
Plugging them into \cref{line:scc:reach1,line:scc:reach2} in \cref{alg:SCC} gives a high-performance parallel SCC algorithm.
We believe that these techniques are general and useful in many graph algorithms.
As proofs-of-concept, in \cref{sec:others}, we apply the proposed ideas to two more algorithms: connected components (CC) and least-element lists (LE-lists), and show new algorithms with better performance. 
\revision{We present notation and parameters used in this paper in \cref{tab:notations}. }

\begin{table}[t]
  \centering
  \small
  \revision{
    \begin{tabular}{cp{7.1cm}}
    \hline
    \multicolumn{2}{l}{\textbf{General notations and vertical granularity control:}}\\
    $n$     & The number of vertices in a graph. \\
    $m$     & The number of edges in a graph. \\
    $P$   & The number of processors available. \\
    $\beta$   & The multiplier of prefix-doubling for SCC, LDD, and LE-List algorithms. Usually $\beta\in(1,2]$. We use $\beta =1.5$ in our system.\\
    $\tau$   & The threshold for vertical granularity control, which is the upper bound of visited neighborhood size per node.
            We use $\tau=512$ as the default value.\\
    \hline
    \multicolumn{2}{l}{\textbf{Hash Bag:}}\\
    $\lambda$ & The first chunk size of \hashbag{}. Theoretically, $\lambda = \Omega((P+\log n)\log n)$.
    We set $\lambda=2^{10}$ in our system. \\
    $\sigma$ &  The threshold of number of samples to trigger \hashbag resizing. Theoretically, $\sigma = \Omega(\log n)$.
    We use $\sigma = 50$ in our system.\\
    \hline
    \end{tabular}%
    \vspace{-1.5em}
  \caption{\small \textbf{Notations used in this paper.}}
  \label{tab:notations}%
  }
\end{table}%

\hide{
Our first technique is a parallel data structure, called \emph{\hashbag},
for maintaining frontiers in graph algorithms.
This is motivated by the challenge of maintaining (sparse) frontiers---to allocate enough memory for the next frontier,
one has to precompute the space needed, which causes extra computation on scanning the edges.
Our goal is to design a space-efficient dynamically-resizable data structure
to avoid processing the frontier multiple times. 
This new data structure is used in all the algorithms in this paper.

The second technique is to overcome the challenge of the high scheduling overhead in
processing large-diameter graphs. Such overhead comes from two aspects.
First of all, on such graphs,
the frontier size and vertex's neighborhood size are usually too small to enable great parallelism.
Secondly, in BFS-based algorithms, the number of rounds is also large,
which causes massive costs on synchronizing between rounds.
Our goal is to allow each parallel task to explore a reasonably large neighborhood for each vertex, possibly larger than one hop.
This greatly reduces the number of rounds needed in the SCC algorithm, and improves the performance.

\myparagraph{Issues in the BFS in \gbbs{} and Overview of Our Solutions.}
The aforementioned BFS algorithm has $O(m)$ work and $\tilde{O}(D)$ span, where $D$ is the diameter of the graph.
It is widely used in practice~\cite{shun2013ligra,gbbs2021}.
However, using it for reachability search is inefficient in two aspects, and we propose solutions to tackle these two challenges.

First, the sparse mode needs to visit every edge twice---the first time counts the size of the next frontier and allocates memory,
and the second time adds the other endpoint to the next frontier.
This doubles the work in multi-reachability searches, and single-reachability searches on large-diameter graphs (in both cases we cannot use the dense mode).
In \cref{sec:hashbag}, we design a space-efficient dynamically-resizable data structure
to avoid processing the frontier multiple times, called the \defn{parallel hash bag}. 
In this way, we only need to visit every edge once when processing the frontier.

Second, the number of rounds in parallel BFS is proportional to the graph diameter.
We need to fork and globally synchronize threads between two rounds, which is expensive in practice.
As a result, parallel BFS performs poorly on large-diameter graphs (e.g., \knn{} and lattice graphs).
Note that what we need in the BGSS algorithm is \emph{reachability, not the shortest-path distance computed by BFS}.
In \cref{sec:local}, we show a novel optimization that can significantly reduce the number of rounds in a reachability search, which greatly improves the performance.

}

\subsection{\titlecap{\vgc}}\label{sec:local}
In this section, we present our \vgc{} (\VGC{}) optimization.
As mentioned, previous work~\cite{gbbs2021} uses parallel BFS for reachability searches,
where the number of rounds is proportional to the diameter of the graph.
On many real-world sparse graphs with large diameters, both the frontier size and the average degree are small, which leads to two challenges.
First, every parallel task (roughly processing one vertex in the frontier) is small,
and the cost of distributing the tasks to the processor can be much more than the actual computation.
Second, the number of rounds is large, resulting in many rounds of distributing and synchronizing threads.

\begin{figure}
  \centering
  \includegraphics[width=\columnwidth]{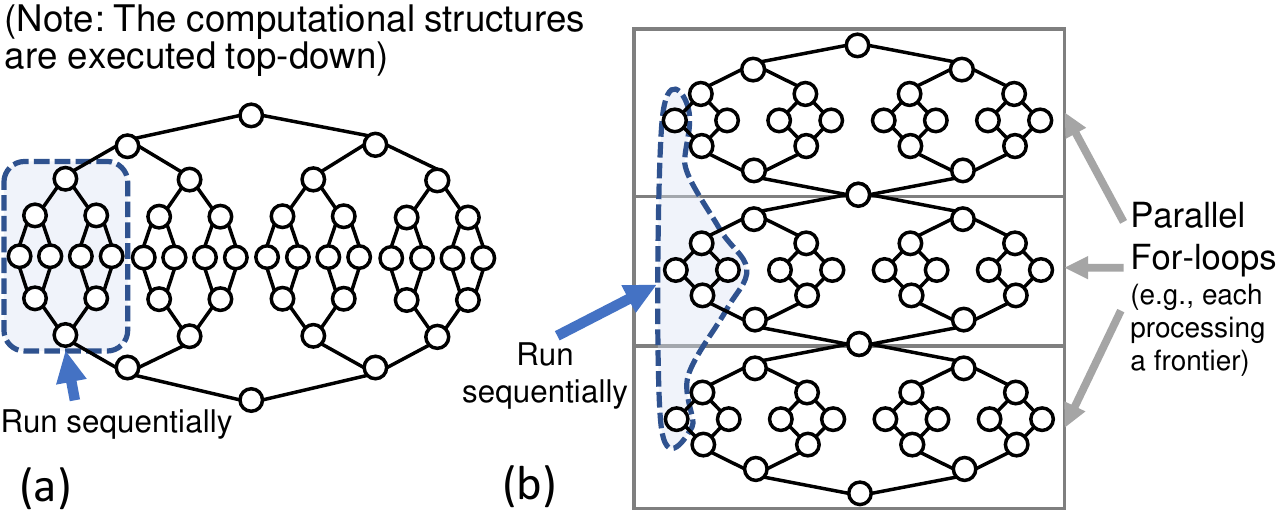}
  \caption{\small \textbf{An illustration of (a) the classic horizontal granularity control (HGC) and (b) our new vertical granularity control (VGC).  } We consider work-stealing schedulers. (a): The computation structure of parallel-for or divide-and-conquer algorithms. HGC groups computation in the same level and run sequentially, in order to reduce scheduling overhead.
  (b): The computation structure of several rounds of parallel-for, each starting from one thread and forking parallel tasks in a nested fashion.
  Each of them can be considered as a round (processing a frontier) in a parallel graph algorithm. VGC groups computations in different rounds and run sequentially, in order to reduce scheduling overhead. It breaks the synchronization points and reduces the number of rounds.
  }\label{fig:gc}
  \vspace{-.25in}
\end{figure} 

Audience familiar with parallel programming must know the concept of \emph{granularity control} (aka.\ coarsening), aiming to avoid the overhead caused by generating unnecessary parallel tasks.
For computations with sufficient parallelism, e.g., a parallel for-loop of size $n\gg p$ where $p$ is the number of processors, most existing parallel software (e.g.,~\cite{acar2019provably,blelloch2020parlaylib,Leiserson2010}) will automatically stop recursively creating parallel tasks at a certain subproblem size and switch to a sequential execution (see \cref{fig:gc}(a)) to hide the scheduling overhead.
We refer to this classic approach as the horizontal granularity control (HCG) since it merges the computation on the same level (sibling leaf nodes in \cref{fig:gc}(a)).

Unfortunately, this idea does not directly apply to reachability searches or similar problems on sparse graphs.
HGC is used when there is excessive parallelism to saturate all processors.
However, when processing sparse graphs, the issue becomes that we have insufficient computation (frontiers with small sizes) to saturate the all processors for good parallelism, and grouping sibling (horizontal) computation in the same round only makes it worse.
To tackle this challenge, we propose a novel and very different approach, referred to as \emph{\vgc{}} (VCG).
The high-level idea is still to increase each task size to hide the scheduling overhead, but we merge the computation across \emph{different levels} to acquire more work and saturate all processors in each round (an example in \cref{fig:gc}(b)).
In this way, we break the synchronization points and reduce scheduling overhead.
Note that this also means that VGC is unlikely to be automatic (unlike HGC)---breaking the synchronization structures may significantly change the computation and needs careful redesign of the algorithm (in our case, we need the new data structure \emph{\hashbag} to deal with non-determinism, see \cref{sec:hashbag}).
In the rest of this section, we will show how to apply VGC to reachability searches and achieve good parallelism. Our motivation is from some theory work for parallel reachability algorithms.


\hide{
For the challenge of reachability queries on large-diameter graphs,
we have the same motivation to make each parallel task sufficiently large to
hide the scheduling cost to distribute the work to different processors.
Despite being straight-forward on computational structures for divide-and-conquer algorithms (stop dividing at a certain level) or simple for-loop (group iterations into sequential blocks), granularity control seems highly non-trivial for graph algorithms since the computation are irregular.
We are not the first to observe such overhead.
Indeed, we are aware of many work~\cite{beamer2015gap,zhang2020optimizing,nguyen2013lightweight,rhostepping} also use specific optimizations to reduce the number of rounds.
However, as the computation in graph algorithms is irregular (unlike divide-and-conquer algorithms),
it is highly non-trivial to do straightforward granularity control.
Many of these optimizations are complicated, or need to deal with low-level thread scheduling.
Most of them are specific for single-BFS or SSSP.
As such, they seem hard to generalize to more complicated problems such as SCC and low-diameter decomposition (LDD, used in CC in \cref{sec:cc}).
In this paper, we formalize this challenge and name the solution as \emph{\vgc{}} (\VGC{}).
Note that the original granularity control works ``horizontally'' because
it merges sibling tasks in the same (horizontal) level (e.g., leaves in divide-and-conquer algorithms) in the computational DAG,
while VGC merges tasks across (vertical) levels (e.g., let a vertex perform work of several rounds without enforcing thread synchronization between them).}

\myparagraph{Motivations from the Theory Work and Our Solution}.
To reduce the number of rounds in BFS-like algorithms, many theoretical results use \emph{shortcuts}~\cite{blelloch2016parallel,Shi1999,ullman1991high,cohen1997using,klein1997randomized,fineman2019nearly,jambulapati2019parallel}
to reduce the diameter of the graph.
Unfortunately, these approaches can be impractical because they incur high overhead for storing the shortcuts, increased memory footprint, and
a significant preprocessing cost (e.g.,
Fineman's algorithm~\cite{fineman2019nearly} has $O(m\log^6 n+n\log^{10} n)$ preprocessing work).
Hence, these algorithms are unlikely to beat the $O(m)$ Tarjan's sequential algorithm using modern computers with tens to hundreds of processors.

\begin{figure}[t]
  \centering
  \includegraphics[width=\columnwidth]{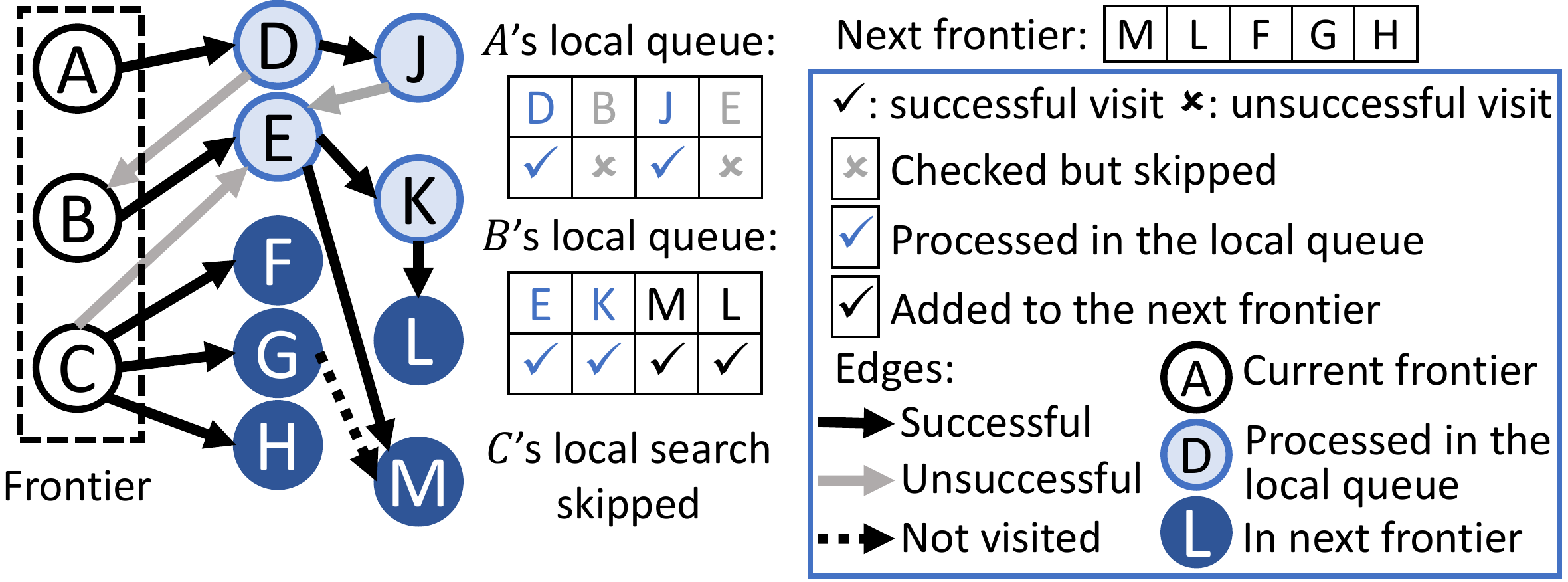}
  \caption{\small \textbf{A possible execution of a local search.} 
  The initial frontier is $\{A,B,C\}$ and $\tau=4$.
  $A$ successfully visits one neighbor $D$. As its local queue is not full, it then visits $D$'s neighbor $B$ (skipped) and $J$.
  $J$ visits $E$ but fails.
  No more vertex are left in $A$'s local queue.
  $B$ visits $E$, and $E$ visits $K$ and $M$.
  Then, $K$ visits $L$ and add it to the queue.
  Now $B$'s queue is full.
  The unfinished vertices $M$ and $L$ will be flushed to the next frontier.
  $C$ has four ($\ge \tau$) neighbors, so we directly check all neighbors and add successful ones ($F$, $G$, and $H$) to the next frontier. 
  We process 2-hop neighbors from the frontier in one round.
  \label{fig:local}}
  \vspace{-.2in}
\end{figure}

To perform \VGC{} without much overhead,
we wish to add shortcuts but avoids explicitly generating them.
Instead of shortcutting \emph{every} vertex~\cite{blelloch2016parallel,Shi1999}, which can be costly, we only shortcut those in the current frontier,
so that a vertex in the frontier can perform some work in the next several rounds ``in advance'' to enable \VGC{}.
Without shortcutting, visiting the same vertices may take multiple rounds to finish.
We wish to process the shortcuts locally and sequentially to avoid space overhead.
For each vertex $v$ being processed, we shortcut it to some (nearest) vertices reachable from $v$ on the fly by a \defn{local search} from $v$, such that we do not need to store the shortcuts.
Our local search is similar to the sequential BFS algorithm. 
We maintain a \defn{local queue} starting from $v$ and visit each of its neighbor $u$.
We will set all sources reachable to $v$ as reachable to $u$.
If any of these sources $s$ is new to $u$, we add $(u,s)$ as a reachable pair, and add $u$ to the tail of the local queue.
We then move to process the next vertex in the local queue.
This process terminates when we have performed ``sufficient work'' in the local search, and we discuss how to control the granularity in \cref{sec:tau}.


We call it the ``local'' queue since we allocate it in the stack memory that is not visible to other processors.
This avoids allocating arrays from the global memory (the heap space), which can be costly (and complicated) in parallel.
The local searches from different vertices in the frontier are independent and in parallel with each other.
The only information that one search needs is whether a vertex has been visited or not, which can be maintained using a boolean array (for single-reachability) or a parallel hash table (for multi-reachability) that support atomic updates.
As such, all searches on the same processor can reuse the same stack space for different local queues.

Using \VGC{}, we can reduce the number of rounds and thus the scheduling overhead since vertices in multiple hops can be visited in one round.
\cref{fig:SCC-rounds} shows that \newchange{\VGC{}} reduces the number of rounds in reachability searches by 3--200$\times$ and greatly improves performance.

Although our idea of using (on-the-fly) shortcuts for \VGC{} is intuitive, two technical challenges remain.
The first is load balancing---vertices can have various degrees and neighborhood patterns, and one vertex may explore a large neighborhood region sequentially.
We discuss the control of granularity (the local search size) in \cref{sec:tau}.
The second is non-determinism.
\hide{
Now each vertex can search out several hops from the frontier.
We can no longer use edge-revisiting to generate the next frontier as is done in previous work (see \cref{sec:prelim}),
because the two rounds of visit may visit different vertices due to runtime non-determinism.
We hence do not exactly know the candidates for the next frontier.}
In \VGC{}, each vertex can search out several hops, and the explored region depends on runtime scheduling.
Hence, we cannot use the edge-revisit scheme in previous work (\cref{sec:bfs}) since the second visit may not perform the same computation as the first one.
To tackle this, we propose the \emph{\hashbag{}} data structure to efficiently maintain the frontier.

\subsection{Control of Granularity}
\label{sec:tau}
The goal to control granularity is to make each task large enough and hide the cost of scheduling it.
However, we cannot let them be arbitrarily large since they are executed sequentially and may cause load-imbalance.
We wish to let all threads perform (roughly) a similar amount of work.
In our implementation, we control the number of visited vertices in each local search by a parameter $\tau$, including both successful and unsuccessful ones.
This number provide an estimation of the workload for each local search.

In particular, when processing a vertex $v$ in the frontier, we first check the number of $v$'s outgoing neighbors.
If it is more than $\tau$, we process all its neighbors in parallel as in the standard way, since we have sufficient work to do and no more shortcuts are needed.
Otherwise, we start the local search and maintain a counter $t$ starting from zero. When processing a vertex $v$ in the local queue,
we increment $t$ for every neighbor visited, successfully or unsuccessfully.
Note that since the local search is performed sequentially,
there is no race condition in maintaining the counters.
We stop the local search either when the queue becomes empty (all possible vertices have been visited),
or when the counter reaches~$\tau$ (this task is reasonably large).
For all remaining vertices in the local queue, we directly add them to the next frontier.
Conceptually, we shortcut $v$ to the $\tau$ nearest vertices that otherwise may need multiple rounds to reach.
We present an illustration in \cref{fig:local}. The desired granularity can be controlled by the parameter $\tau$.

\revision{Intuitively, we can choose the parameter $\tau$ empirically based on traditional HGC base-case size:
usually using base-case size around 1000 operations is sufficient to hide scheduling overheads.}
We also experimentally study the value of $\tau$ in \cref{sec:exp:tech}.
Compared to plain BFS (no \VGC{} used),
we noticed that the performance on most graphs (both large- and low-diameter graphs) except three graphs improved using any $1<\tau \le 2^{16}$.
Overall, the performance is not sensitive in a large parameter space $2^6\le \tau\le 2^{12}$ on almost all graphs.
We simply set $\tau=2^9$ as the default value, \revision{which is similar to typical HGC threshold}.
One can control granularity using other measures such as the number of generated reachability pairs or successfully visited vertices.
We believe that the measure of granularity is independent with the idea of \VGC{}.
We plan to explore more criteria to control granularity for \VGC{} in future work.



\subsection{Parallel Hash Bag}\label{sec:hashbag}

As mentioned, \VGC{} brings up the challenge in maintaining the frontier efficiently.
Recall that in parallel BFS, the ``edge-revisit'' scheme first visits all edges incident the frontier to decide the successfully visited vertices,
and then revisits all edges to output them to a consecutive array as the next frontier.
Since the candidate of the next frontier $\ff_{i+1}$ are all neighbors of $\ff_i$, we can use a boolean flag to record the success information and let the second visit do the same computation as the first time.
However, with \VGC{}, each vertex can search out several hops, and the order of the searches is decided by the runtime scheduling.
Note that the local queue is stored in the stack space and discarded after the search.
If we want to borrow the ``edge-revisit'' scheme in BFS, we need to explicitly store the information of the local queues, which can be very costly.
To tackle this challenge, we propose a new data structure called \defn{parallel \hashbag{}} to maintain the frontier efficiently, such that the next frontier can be generated by visiting the edges only once.
Our \hashbag{} supports \bagput{}, \revision{\bagpack{}}, and \bagforall{} efficiently both in theory  (work, span, and I/O) and in practice.
\hide{
Our design is based on two observations.
First of all, in most of the applications, the functionality needed is to put an element into the set
and to process all elements (possibly in parallel) in the set, whereas a fast look up operation is not required.
Therefore, we borrow the idea from some existing papers and call the needed data structure a \defn{bag}.
Secondly, to maintain a reasonable size of the bag, it needs to be dynamically resizable while concurrent insertions going on. However, we wish to avoid
memory allocation or copying.}
We start with defining the interface of \hashbag{s}:
\begin{itemize}[leftmargin=*,topsep=0pt, partopsep=0pt,itemsep=0pt,parsep=0pt]
  \item \bagput{}($v$): add the element $v$ into the bag (resize if needed). \revision{It can be called concurrently by different threads.}
  \revision{
  \item \bagpack{}(): extract all elements in the bag into an array and remove them from the bag.
  }
  \item \bagforall{}(): apply a function to all elements in the bag in parallel.
\end{itemize}

In \hashbag{s}, we require to know an upper bound $n$ of the total size,
which is true for most applications of \hashbag{s} (e.g., $n=|V|$ for maintaining frontiers).
\revision{We pre-allocate $O(n)$ number of slots as an array $\bagarray$ to hold elements to be inserted.
However, instead of directly using all the slots, we only use a prefix of them with $O(s)$ in expectation,
where $s$ is the current number of elements in the bag.
}
This guarantees the efficiency for \bagpack{} and \bagforall{} since we only need to touch $O(s)$ space to process $s$ elements.
The problem then boils down to maintaining the right size of the used prefix and how to ``resize'' efficiently.

\hide{
\begin{algorithm}[t]

\caption{Struct of Hashbag\label{alg:Biconnectivity}}
\SetKwProg{myfunc}{Function}{}{}
\SetKwFor{parForEach}{ParallelForEach}{do}{endfor}
\SetKwProg{mystruct}{Struct}{}{}

  Array \bagarray{}$[]$ of data type\\
  Array $t[]$ \tcp*{tail of chunk $i$}
  Array $s[]$ \tcp*{\# samples taken in chunk $i$}
  int $r\gets 0$ \tcp*{the current round}
  constant $0<\alpha<1$ \tcp*{desired load factor}
  \myfunc{\upshape \fname{Hashbag}($n$)\tcp*[f]{Initialization}} {
    allocate \bagarray{} of size $n/\alpha$\\
    allocate $t[]$ and $s[]$ of size $\log n/\alpha$\\
    $s[i]\gets 0$ for all $i$;\\
    $t[0] \gets 0$\tcp*{pre-set tails for each chunk below}
    $t[1] \gets n^\lambda$;\\
    \For {$i\gets 1$ to $\log n/\alpha$ } {
      $t[i+1]\gets t[i]\times 2$\\
      \lIf {$t[i+1]> n/\alpha$} {$t[i+1]\gets n/\alpha$}
    }
  }
  \myfunc{\upshape \bagput($k$)} {
    $r^*\gets r$;\\
    \If {sampled successfully} {
      \faa$(\&s[r^*])$;\\
      \If {$(s[r^*])$ shows the current chunk is too full} {
        \bagresize$(r^*)$;\\
        \Return {\bagput$(k)$}
      }
    }
    $i\gets$ a random position in $t[r^*-1]..t[r^*]$\\
    \While {$\neg$\cas(\&\bagarray$[i]$, \fname{empty}, $k$)}{
      $i\gets i+1$\\
      \lIf {$i=t[r^*]$}{$i\gets t[r^*-1]$}
      \If {has probed more than $\kappa$ times}{
        \bagresize$(r^*)$;\\
        \Return {\bagput$(k)$}
      }
    }
  }
  \myfunc{\upshape \bagresize($r^*$)} {
    \cas$(r,r^*,r^*+1)$;
  }
  \myfunc{\upshape \bagfind($k$)} {
    $r^*\gets r$\\
    \While{$k$ not found in chunk $r^*$}{
      $r^*\gets r^*-1$;\\
      \If {$r^*<0$} {\Return{$\false$}}
    }
    \Return{$\true$}
  }
  \myfunc{\upshape \bagforall$(f)$ \tcp*[f]{Apply function $f$}} {
    \parForEach{$x\in \bagarray$} {
      \lIf {$x$ is not empty} { $f(x)$ }
    }
  }
  \myfunc{\upshape \bagpack$()$} {
    Parallel pack for non-empty elements in array \bagarray{};
  }
\end{algorithm}
}

\begin{figure}
\centering
\input{code.tex}
\begin{lstlisting}[frame=lines]
struct Hashbag {
  double $\alpha$; @\hfill@// desired load factor
  data* bag;
  int* tail;@\hfill@// tail of chunk $i$
  int* sample;@\hfill@// #samples of chunk $i$
  int r = 0; @\hfill@// current chunk id  @\codegap@
  Hashbag(int n) { @\hfill@// constructor
    allocate tail[] and sample[] of size $\lceil\log \frac{n+\lambda}{\alpha}\rceil$;
    initialize sample[] and tail[] to 0;
    tail[0] = $\lambda$;@\label{line:gettail1}@
    for (i = 1; i< $\lceil\log \frac{n+\lambda}{\alpha}\rceil$; i++) tail[i] = tail[i-1]*2;@\label{line:gettail2}@
    bag = new data[tail[i-1]];}@\codegap@
  void insert(data k) {
    r@'@ = r;
    if (sampled successfully) {
      // Lines @\ref{line:faastart}@-@\ref{line:faaend}@: @\faa{}@ using @\cas@
      t = sample[r@'@];@\label{line:faastart}@
      while (!@\cas@(&sample[r@'@], t, t+1)@\label{line:bagcas}@@\label{line:faaend}@) {
      if (sample[r@'@] too large) {  @\hfill@// current chunk is full
        try_resize(r@'@);
        return insert(k); } } }
    i = random position in tail[r@'@-1]..tail[r@'@]@\label{line:putselectrandom}@
    while (!@\cas@(&bag[i],$\emptyset$,k)@\label{line:putcas}@) {
      i = next(i);@\label{line:putprobe}@@\hfill@// linear probe to find the next slot
      if (has probed more than $\kappa$ times) {
        try_resize(r@'@);
        return insert(k);  }  }  }@\codegap@
  void try_resize(int r@'@) {@\cas@(&r, r@'@,r@'@+1);}}
\end{lstlisting}
\caption{\small \textbf{Pseudocode for the \hashbag{}. }\label{fig:algo:hashbag}}
\vspace{-0.4cm}
\end{figure} 

We show (part of) the pseudocode of \hashbag{s} in \cref{fig:algo:hashbag} and an illustration in \cref{fig:hashbag}.
The size of the $\bagarray$ is preset as $\Theta(n/\alpha)$,
\revision{where $\alpha$ is the desired load factor and $n$ is the upper bound of total size as mentioned before}.
\revision{We conceptually divide the $\bagarray$ into chunks and use them one by one.
A resizing means moving to the next chunk for use.
The chunks have doubling sizes of $\lambda, 2\lambda, 4\lambda$ ..., where $\lambda$ is a parameter for the first chunk size.
At initialization, we set up an array $\tail{}[\cdot]$ , where $\tail{}[i]$ is the end index of the $i$-th chunk.
We use a variable $r$ to indicate the current in-use chunk id, starting from $0$.
Elements are always inserted into the $r$-th chunk (indices from $\tail[r-1]$ to $\tail[r]$ for $r\geq 1$).
}
\revision{

An \bagput{} randomly selects an empty slot in this chunk (Line \ref{line:putselectrandom}),
attempts to put the element in this slot using \CAS{},
and linear probes if the \CAS{} fails (Lines \ref{line:putcas}--\ref{line:putprobe}).
Note that different from the hash table, the \bagput{} on \hashbag{s} does not check duplicates, but all applications in our paper (maintaining frontiers)
can ensure that no duplicates will be added to the bag.
For example, duplicates can be checked before calling \bagput{}, e.g.,
using a boolean flag for each vertex to indicate if it is in the frontier (the array $\vname{visit}$ in \cref{alg:BFS_hashbag},
details explained below).
}

\newcommand{\ssize}{\sigma}

\begin{figure}
  \centering
  \vspace{-.1in}
  \includegraphics[width=\columnwidth]{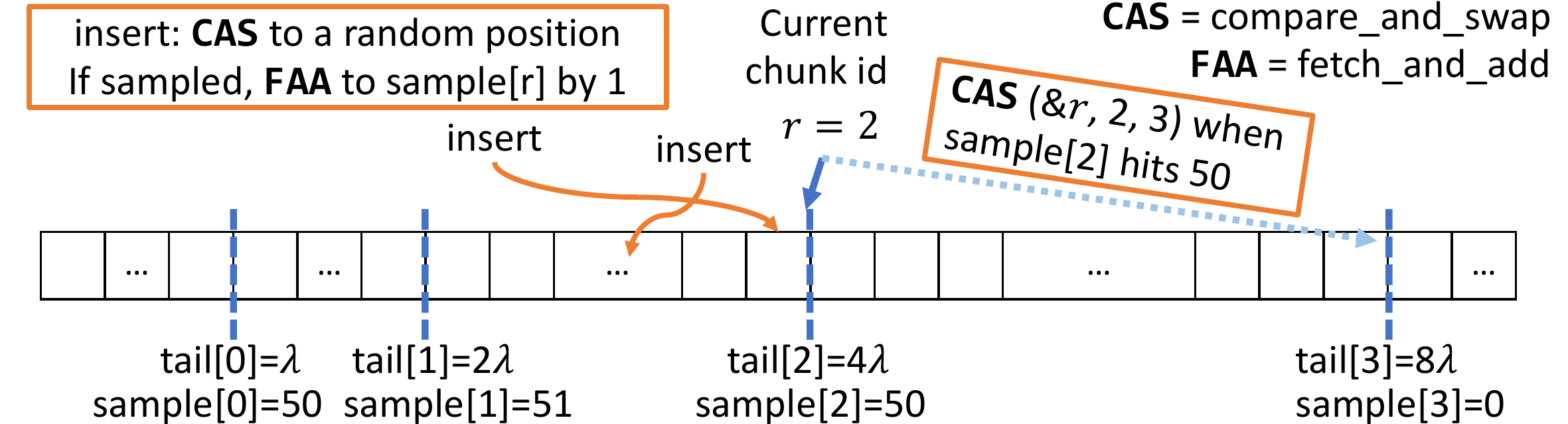}
  \vspace{-.1in}
  \caption{\small \textbf{Parallel Hash Bag.}
  The \hashbag{} is a preallocated array of size $O(n)$, split into chunks of exponentially grown sizes, starting from $\lambda$.
  $\tail[i]$ is the last index to use for chunk $i$.
  The current chunk id is $r$. An \bagput{} puts the element to a random position in the current chunk (linear probe for conflict/collision).
  Each element is sampled at a certain rate. $\sample[i]$ is the number of samples in chunk $i$. When the $\sample[r]$ reaches a threshold ($\ssize=50$ in this example), 
  we resize it by \CAS{} $r$ to $r+1$. 
  }\label{fig:hashbag}
  \vspace{-.2in}
\end{figure} 

To efficiently decide when resizing is needed, we use a sampling strategy to estimate the size of the \hashbag{}.
We use $\sample{}[\cdot]$ to count the number of samples in chunk $i$, and resize when the number of samples hits $\ssize$.
\revision{We fix $\sigma$ for all chunks, but set sample rate accordingly for each chunk as $\ssize/\alpha$ divided by the chunk size.
Conceptually, this means to trigger a resize once the load factor goes beyond $\alpha$.
The larger the chunk is, the smaller the sampling rate is.
Theoretically, getting accurate estimations 
requires $\ssize=\Omega(\log n)$.
}

In \bagput{}, we sample the element with the current rate. 
If sampled successfully, we increment $\sample[r]$ ($r$ is the current chunk) by 1 atomically by \CAS{} (conceptually this is an atomic \faa{} operation).
When $\sample[i]$ hits $\ssize$, a constant fraction of this chunk is full \whp{}, so a resizing attempt is triggered (\texttt{try\_resize}).
Also, when we linear probe for more than a certain number of times, we also trigger a resizing (although this should be rare).
\revision{
In both cases, we resize by increasing $r$ by $1$ using a \CAS{}, and call \bagput{} again to add this element to the new chunk.
}

  
\bagpack{} and \bagforall{} are applied to all elements in $\bagarray[\cdot]$ up to the current chunk $r$ (indices from $0$ to $tail[r]$).
\bagpack{} uses a standard parallel pack~\cite{JaJa92} to output all (non-empty) elements in an array and remove them from the bag in parallel.
\bagforall{} calls a parallel for-loop to apply the function on all elements (skip the empty slots).
\hide{
For \bagforall{}, we can use a parallel for-loop on all elements in $\bagarray[\cdot]$ up to chunk $r$ and skip the empty ones.
A regular parallel pack~\cite{JaJa92} can be used for \bagpack{}.
}

\revision{We present the framework on using a \hashbag{} $H$ for reachability query in \cref{alg:BFS_hashbag}.
Given the current frontier $\ff_i$, we will
visit all vertices in $\ff_i$ in parallel and perform local searches from them.
We use an array of boolean flags $\vname{visit}[\cdot]$ to record whether each vertex has been visited.
When a vertex $v\in \ff_i$ visits a vertex $u$,
we will use \CAS{} to set $\vname{visit}[u]$ as \true{} (\cref{line:bfs:cas}).
As mentioned, \CAS{} guarantees that only one concurrent visit to $u$ will succeed.
Note that if $\vname{visit[u]}$ is already $\true$,
this if-condition will also fail, which guarantees no duplicates in the \hashbag{}.
If the \CAS{} succeeds, we call $\bagput(u)$ to add $u$ to the \hashbag{}.
Note that if local search is enabled, vertices visited within the local search are not added to the next frontier (see details in \cref{sec:local}).
We omit such cases in the pseudocode for simplicity.
Finally, when we finish exploring all the vertices in $\ff_i$, we extract (emit and clean) all vertices from the \hashbag{} to form the next frontier (\cref{line:bfs:pack}).
}

\myparagraph{Theoretical Analysis of Hash Bags.} We now show the cost bounds of the \hashbag{}.

\vspace{-.05in}
\begin{theorem}\label{thm:hashbag}
  For a parallel \hashbag of total size $n$ and first chunk size $\lambda=\Omega((P+\log n)\log n)$, inserting $s$ elements using $P$ processors costs $O(s)$ expected work and $O(\log s\log n)$ span \whp{}, and listing or packing $s$ elements uses $O(s+\lambda)$ work and $O(\log s)$ span, both \whp{}, with mild assumptions (see below).
\end{theorem}
\vspace{-.05in}
We provide the formal proof in the full in \cref{app:hashbagsize}.
In the analysis, we assume the threads are loosely synchronized,
where between two consecutive executions of Line~\ref{line:bagcas}, other processors can execute at most a constant number of instructions.
This assumption is reasonable in practice and is used in analyzing other parallel algorithms such as the analysis of the work-stealing scheduler~\cite{Acar02,gu2022analysis}.
Note that the value $P$ is usually a small number (up to hundreds) in practice,
and can generally be considered as polylogarithmic of input size $n$.
In practice, we set $\lambda=2^{10}$ and $\ssize=50$.
We pick $\ssize=50$ since it is close to $\log n$.
\revision{We use $\lambda=2^{10}$ since our analysis indicates $\lambda$ should roughly be $\log^2 n$.
We tested $\lambda$ for a large range and it affected the running time minimally for $2^8\le \lambda\le 2^{16}$,
so we simply use a single value for all tests.}

Our experiments show that \hashbag{s} are fast in practice due to the space efficiency and fewer memory accesses.
Although we design \hashbag{s} for \VGC{}, our experiments show that \hashbag{} itself also improves the algorithms' performance
because it avoids scanning the frontier twice. When applying it to LE-lists (see \cref{sec:lelists}), where we can use \hashbag{s} but not \VGC{}, we also achieve up to 10$\times$ speedup over existing implementations.

\hide{
\myparagraph{Background.} Like the BFS in \cref{sec:bfs}, many graph algorithms are based on processing \emph{frontiers} of vertices~\cite{shun2013ligra,gbbs2021}.
Since the frontier is generated dynamically during the algorithm,
its size can vary significantly between rounds and is unpredictable. 
Existing libraries (e.g., Ligra~\cite{shun2013ligra} and GBBS~\cite{gbbs2021})
tackle this by precomputing the size of the next frontier as described in \cref{sec:bfs}.
This requires scanning the edges \emph{at least twice},
which almost doubles the cost since the I/O of scanning edges is the performance bottleneck for most graph algorithms.
The second solution is to allocate memory using an upper bound.
For instance, the SCC algorithm in GBBS~\cite{gbbs} stores all reachable vertex pairs
using phase-concurrent hash tables~\cite{shun2014phase}.
Since resizing is not allowed during insertions, the hash table size has to be an upper bound of the needed size, which can be very loose.
Our experiment shows that the average load factor of their hash table is lower than 20\% (see Section \ref{sec:exp:loadfactor}).
This incurs overhead both in space and time due to increased memory footprint.
Using dense mode avoids the problem.
However, on large-diameter graphs
and for the multi-source search in the SCC algorithm,
we can hardly take advantage of dense modes.
}

\SetInd{0.5em}{0.7em}
\begin{algorithm}[t]
  \small
  \revision{
    \caption{{Parallel Single-Reachability Using Hash Bags\label{alg:BFS_hashbag}}}
    \KwIn{A directed graph $G=(V,E)$ and a set sources $S\in V$}
    \SetKwFor{parForEach}{parallel\_for\_each}{do}{endfor}
    \SetKwInOut{Maintains}{Maintains}
    \DontPrintSemicolon
    $\ff_0=S$\\
    $\vname{visit}[v]\gets \false $ for all $v\in V$ except $\vname{visit}[s]\gets true$ \\
    $i\gets 0$\\
    $H \gets \mathit{Hashbag()}$ \Comment{initial $H$ as an empty \hashbag{}}\\
    \While {$\ff_{i}\ne \emptyset$}{
    \parForEach{ $v\in \ff_i$}{
        Visit $v$'s neighborhood, use local search if applicable\\
        \ForEach(\Comment{Processing a reachability pair $(v,u)$}){$u$ visited by $v$}{
          \If( {(*)} ) {\cas$(\&\vname{visit}[u], \false, \true)$ \label{line:bfs:cas}} {
                $H.\bagput{}(u)$    \label{line:bfs:insert}
              }
        }
    }
    $\ff_{i+1} \gets H.\bagpack{}()$ \Comment{pack elements and clean the bag} \label{line:bfs:pack}\\
    $i\gets i+1$
    }
    {(*)}: Note: vertices visited within local searches will not be added
  }
\end{algorithm}

\hide{
\begin{algorithm}[t]
  \small
  \revision{
    \caption{{Parallel Single-Reachability using \hashbag{}\label{alg:BFS_hashbag}}}
    \KwIn{A directed graph $G=(V,E)$ and a source $s\in V$}
    \SetKwFor{parForEach}{parallel\_for\_each}{do}{endfor}
    \SetKwInOut{Maintains}{Maintains}
    \DontPrintSemicolon
    $\ff_0=\{s\}$\\
    $\vname{visit}[v]\gets \false $ for all $v\in V$ except $\vname{visit}[s]\gets true$ \\
    $i\gets 0$\\
    $H \gets \mathit{Hashbag()}$ \Comment{initial $H$ as an empty \hashbag{}}\\
    \While {$\ff_{i}\ne \emptyset$}{
    \parForEach{ $v\in \ff_i$}{
        $Q \gets queue(\tau)$ \Comment{initial local queue $Q$ with length $\tau$}\\
        $Q.put(v)$\\
        $cnt \gets 1$ \Comment{initial the number of vertices visited as $1$} \label{line:bfs:cnt}\\
        \While{$cnt < \tau$}{
          $u = Q.pop()$ \\
          \If {$u$ has more than $\tau$ neighbors}{
            \parForEach{unvisited neighbor $w$ of $u$ \label{line:bfs:start}}{
              \lIf {$\CAS{}(\&\vname{visit}[w], \false, \true)$ \label{line:bfs:cas}} {
                $H.\bagput{}(w)$    \label{line:bfs:insert}
              }
            }
            \textbf{break}\\
          }\Else{
            \For {unvisited neighbor $w$ of $u$}{   \label{line:local:start}
              $cnt \gets cnt+1$\\
              \lIf {$\CAS{}(\&\vname{visit}[w], \false, \true)$ \label{line:local:cas}} {
                $Q.put(w)$        \label{line:local:insert}
              }
            }
          }
        }
        \lFor(\Comment{add to the next frontier}){$u$ in $Q$}{
          $H.\bagput{}(u)$                          \label{line:local:end}
        }
    }
    $\ff_{i+1} \gets H.\bagpack{}()$ \Comment{pack elements and clean the bag} \label{line:bfs:pack}\\
    $i\gets i+1$
    }
  }
\end{algorithm}
}

\hide{
We start with a prefix of size $\lambda$, and double the size every time of resizing.
We use an array $\tail{}[\cdot]$, where $\tail{}[i]$ is the tail index after the $i$-th resizing,
and set $\tail{}[i]$ at initialization (Lines \ref{line:gettail1}--\ref{line:gettail2}).
Each resizing adds a chunk of the array at the end.
The current chunk being used is noted by a variable $r$, starting from $0$.
Elements are always inserted into the last chunk (indices from $\tail[r-1]$ to $\tail[r]$).
An \bagput{} randomly selects a position in this chunk (Line \ref{line:putselectrandom}), attempts to put the element in this position
using \CAS{}, and linear probes if the \CAS{} fails (Lines \ref{line:putcas}--\ref{line:putprobe}).
}

\hide{
A resizing attempt is done through a \cas{} on the current $r$ to be $r+1$. Note that only one such concurrent resizing will succeed.
No matter if the resizing (the \cas{}) is successful or not, a new chunk must have been added. This is also why we need to call \bagput{} again to grab the new value of $r$ and redo \bagput{}.
As we know an upper bound on the number of elements that will be put into the \hashbag{} is $n$, the size of the array will grow to no more than $\Theta(n)$. Therefore, we never need to reallocate new memory or move the elements to another memory chunk.

} 
\section{Implementation Details}\label{sec:implementation}

We use the techniques in~\cref{sec:technique} (\VGC{} with hash bags) to implement reachability searches in the BGSS algorithm for SCC (\cref{alg:SCC}).
This section further presents some details in the implementation.
Many of these ideas are also adopted in other recent parallel SCC implementations or graph libraries~\cite{gbbs2021,ji2018ispan,slota2014bfs,hong2013fast,mclendon2005finding}.
We summarize the cost of our implementation in five categories: \defn{trimming}, \defn{first SCC}, \defn{multi-search}, \defn{labeling}, and \defn{hash table resizing}.
In \cref{sec:exp:breakdown}, we show a running time breakdown based on these five categories.

\myparagraph{\ref{sec:implementation}.1~\underline{Trimming}.} The algorithm first filters all vertices with zero in- or out-degrees, since they must be in isolated SCCs.
It is used in almost all existing SCC implementations.

\myparagraph{\ref{sec:implementation}.2~Finding the \underline{first SCC}.} 
As the first reachability search in BGSS only contains one source, we use single-reachability to find the first SCC,
and use the standard \emph{dense-backward}~\cite{shun2013ligra,Beamer12} optimization.
This optimization is designed for single-BFS when the frontier is large.
Instead of checking all the \emph{out-}edges from $\ff_{i}$,
the dense mode checks each unvisited vertex $u$ and its \emph{in-}neighbors.
If any of $u$'s in-neighbor is in the previous frontier $\ff_i$, $u$ must be reachable from the source, and we can skip the rest of the edges incident $u$ to save work.
We refer to this optimization as \defn{dense mode}, and the aforementioned approach as \defn{sparse mode}.
We note that dense mode does not work in multi-reachability searches---even if we find a neighbor of $u$ in $\ff_i$, we cannot skip the rest of the neighbors since they may come from different sources than $v$.
Therefore, we only use dense mode in single-reachability searches.

\myparagraph{\ref{sec:implementation}.3~\underline{Multi-}reachability \underline{search}.} Next, we start $(\log n)-1$ batches of multi-reachability searches in both forward and backward directions, where round $i$ uses $2^{i}$ sources (Lines \ref{line:scc:reach1}--\ref{line:scc:reach2}).
During the multi-reachability search, we need a hash table to identify the duplicate reachability pairs.
We use the phase-concurrent hash table~\cite{shun2014phase}.
To avoid high overhead in hash table resizing, we use a heuristic to estimate the hash table size, which is discussed below in Sec.~\ref{sec:implementation}.5.

\myparagraph{\ref{sec:implementation}.4~\underline{Labeling}.} After finding all reachability pairs, we mark all vertices strongly connected with any source as \emph{finished}, and label them using the largest vertex id in this SCC (\cref{line:scc:label1}).
For the other vertices, we need to compute their ``signatures'' of reachability to determine cross edges.
We do this also by setting a label for them (\cref{line:scc:label2}),
which is a hash value of the set of vertices reachable from and to $v$ (combining $R_1$, $R_2$ with its current label).
In this way, two vertices with different labels are in different SCCs.
We set the hash value also as the largest vertex id among all vertex reachable from or to $v$.
To avoid the cost of explicitly removing the cross edges,
we just skip cross edges in later reachability searches if the endpoints have different labels.

\myparagraph{\ref{sec:implementation}.5~Heuristic for \underline{hash table resizing}.} 
The phase-concurrent hash table~\cite{shun2014phase} requires knowing the upper bound of the size before concurrent insertions.
With \VGC{}, we do not know a tight upper bound of the number of reachability pairs $(v,s)$, 
since $v$ can be several hops away from $s$, and the number of possible pairs can be large.
Instead, we compute the number of pairs $a$ in the previous frontier and the number of unfinished vertices $b$, and use $\max(0.3b,1.5a)$ and round it up to the next power of 2 as the next hash table size.
We resize our hash table once an insertion probes too many times. 
This heuristic is inspired by some recent analyses of the BGSS algorithm in~\cite{blelloch2021read}.
As shown in \cref{fig:SCC_Total_Break}, on many graphs, resizing hash tables can be costly, and our heuristic effectively reduces this cost.



\hide{
We start from the strongly connected components (SCC) problem.  
Our implementation is based on BGSS SCC algorithm~\cite{blelloch2016parallelism}
since the algorithm is reachability-based and has theoretical guarantees.
By plugging in our new techniques, our version is significantly faster than
the existing implementation of this algorithm in GBBS~\cite{gbbs2021} (6x faster on average across 18 graphs).

Given a directed graph $G=(V,E)$, a \defn{strongly connected component (SCC)} is a maximal subset $C\subseteq V$ such that for every $u,v\in C$, there are directed
paths both from $u$ to $v$ and from $v$ to $u$.


\myparagraph{Implementation in GBBS~\cite{gbbs2021}}.
Dhulipala et al.~\cite{gbbs2021} engineered BGSS algorithm as part of the GBBS library~\cite{gbbs2021}.
There are two SCC algorithms in GBBS, and we refer to the \emph{RandomGreedy} algorithm as this version is the faster in most of our tests.
The high-level idea is given in \cref{alg:SCC} (the same as ours).
The algorithm first trims all vertices with zero in- or out-degrees and marks them as isolated SCCs (\cref{line:trim}).
Then the prefix-doubling scheme partitions the vertices into batches of size $1, 2, 4, 8, \cdots$ (\cref{line:scc_partition}).
The first batch (one vertex) requires two single-reachability searches to compute the first SCC and cut the edges (Lines \ref{line:firstscc1}--\ref{line:firstscc2}).
The other batches require multi-reachability queries (Lines \ref{line:scc:reach1}--\ref{line:scc:reach2}) that compute a set of \emph{reachable pairs} $(s,u)$ where source $s$ is reachable from/to vertex $u$.
To record the reachable pairs in multi-reachability searches, GBBS uses the phase-concurrent hash table~\cite{shun2014phase}.
The reachable pairs are used to assign labels to the vertices,
and will be used to identify edges \newchange{connecting two vertices with different reachability information} (Lines \ref{line:scc:reach3}--\ref{line:scc:reach4}).
In GBBS, both the single- and the multi-reachability queries are based on BFS, which can be slow on large-diameter graphs since BFS needs too many rounds.
Also, to generate the frontiers, their BFS visits every edge multiple times (see \cref{sec:prelim} for details), which is inefficient even on low-diameter graphs,
as the dense mode is not applicable to multi-reachability searches.

}

\hide{
The major part of BGSS algorithm is the $2n$ reachability searches---the first two run on its own, while the rest of them are grouped into $2\log n$ groups.
GBBS uses parallel BFS for reachability queries, which only works well on small-diameter graphs.
For the multi-reachability searches (after the first round), to check if a certain search from $u$ has already visited a vertex $v$, Dhulipala et al.~\cite{gbbs2021} uses the parallel hash table from~\cite{shun2014phase} that does not support resizing during insertion.
Hence, they allocated a sufficiently large hash table, which introduces overhead on listing the vertices, which is used in cutting the edges. \letong{Overhead also comes from estimate tabel size every round and resizing. }
}




\hide{

\textbf{Implementation Background} If the in- or out-degree of a vertex is zero, then it can not form SCC with other vertices by the definition.  Trimming these nodes at the beginning can make the algorithm run faster.
Multistep \cite{slota2014bfs} runs four steps in order: trimming, one round FW-BW searching, Coloring and serial Tarjan's algorithm.  It runs fast on certain social-network graphs that has one large SCC containing the majority of the vertices, but it costs $O(n^2)$ work in the worst case. ISPAN \cite{ji2018ispan} also takes advantage of trimming and FW-BW searching, but it implements the reachability queries by spanning-tree. ISAN can outperform BFS and DFS based algorithms, but still doesn't have theoretical bounds.
\citeauthor{gbbs2021} presents the first implementation of the BGSS\cite{blelloch2016parallelism} SCC algorithm in \cite{gbbs2021}. It has $O(m \log n)$ expected work and $O(diam(G) \log n)$ depth w.h.p. on the PW-TRAM. The pseudo code is in Algorithm \ref{alg:SCC}. Our implementation also base on it.

The \mf{Partition($V$)} in Line \ref{lst:line:Partition} permutes the vertices in $V$ in random order, and partition them into $\log n$ batches.  $L$ stores each vertex SCC label, which is initialized to be $\infty$. $Done$ records whether vertices are already assigned to a known SCC, and it is set to $0$ at the beginning.

\textbf{Optimizations}

\mf{MarkReachable} is a batching reachability search. \citeauthor{gbbs2021} implement it in a BFS manner: begin with a subset of vertices (frontier), calling edgeMap on it and generate new frontier until not frontier size is zero. They use two resizable hash tables storing $(vertex, label)$ pair to record the reachability information. The size of resizable table is overestimated before each BFS rounds. When resizing, the elements in the old table are reinserted into a new table.  We implement \mf{MarkReachable} in a reachability manner, which allow local search. The benefits of local search is that it can reduce the number of rounds of edgeMap, which can benefit from the fewer synchronizations between rounds.

Once the reachability queries are done both on $G$ and $G^T$, \citeauthor{gbbs2021} take the union of the two hash table by looking for one hash table's element in the other table. The union of the two table forms an SCC, the algorithm mark the vertices in the union as done.  For the vertices not in the union of the two table, it will be assigned with the maximum label that can reach it.

Beside the algorithm shown in \ref{alg:SCC}, both \gbbs{} \cite{gbbs2021} and our implementation trim at the beginning, and apply forward and backward reachability from a single source for the first round.}


%
\hide{Besides, it can be proved that SCCs are completely inside one of the remaining subsets, but not cross them \cite{coppersmith2003divide}.
Thence the algorithm recurses on the remaining subsets. \citeauthor{coppersmith2003divide} show if pivots are picked randomly, this algorithm sequentially runs in $O(m\log n)$ in expectation. \cite{coppersmith2003divide}

The algorithm in \cite{coppersmith2003divide} does not have a guarantee on depth, because the subproblems may be unbalanced. In the case that all vertices are disconnected, most of the vertices will fall into the set that is not reachable from the pivot, which has $\Theta(n)$ recursive depth.  \citeauthor{schudy2008finding} proposed a parallel algorithm with $O(\log ^2n)$ reachability queries depth but cost $O(\log n)$ factor of extra work \cite{schudy2008finding}.  \citeauthor{blelloch2016parallelism} show that if reachability queries are applied in random order, the cost bound $O(n\log n)$ still holds. Besides, queries can be batched into $O(\log n)$ exponentially increasing batches, where queries within each batch can be processed at the same time. This idea is called prefix-doubling. The BGSS SCC algorithm in \cite{blelloch2016parallelism} has $O(W_R(n,m)\log n)$ expected work and has $O((D_R(n,m)+\log n)\log n)$ depth $whp$ on the CRCW PRAM.
}

\section{Other Relevant Algorithms}\label{sec:others}

The two general techniques (VGC and parallel hash bag) introduced in \cref{sec:technique} are general. In this section, we use them to accelerate two other graph algorithms.
In particular, in \cref{sec:cc} we show how to apply these techniques in a parallel graph connectivity algorithm, and in \cref{sec:lelists} we show that using parallel hash bag in the algorithm for least-element lists can lead to significantly faster performance.

\begin{algorithm}[t]
    \caption{\small LDD-UF-JTB Algorithm for Connectivity~\cite{SDB14}}
    \label{algo:connectivity}\small
    \SetKwFor{parForEach}{parallel\_for\_each}{do}{endfor}
    \SetKwProg{myfunc}{Function}{}{}
    \KwIn{A graph $G=(V,E)$  with $V = \{v_1,\ldots,v_n\}$}
    \KwOut{The connectivity labels $L(\cdot)$ of $V$}
    \DontPrintSemicolon
        \smallskip
        $L \leftarrow \fname{LDD}(G)$ \label{algo:connectivity:ldd}\\
        \parForEach{\upshape $(v, u) \in E$} {
            \lIf{$\fname{Find}(L(v)) \neq \fname{Find}(L(u))$} {
                \fname{Union}($L(v), L(u)$) \label{algo:connectivity:union}
            }
        }
        \Return{$L(\cdot)$}

        \myfunc{\upshape \fname{LDD}$(G=(V,E))$} {
            Set $\vname{visit}[v] \leftarrow \false$ and $L(v) \leftarrow v$ for all $v \in V$\\
            $B \leftarrow$ Permute $V$ and group vertices into $O(\log n)$ batches in exponentially increasing sizes \label{algo:ldd:batches}\\
            $F \leftarrow B_1$  \\
            Set $\vname{visit}[v] \leftarrow \true$ for all $v \in B_1$ \\
            \For{\upshape $i \gets 2, \dots , |B|$\label{line:ldd:loop}} {
                $F' \leftarrow \emptyset$\\
                \parForEach{\upshape $v \in F$} { \label{algo:ldd:bfs_start}
                    \parForEach{\upshape $u: (u,v)\in E,\vname{visit}[u] = \false$} {
                            $\vname{visit}[u] \leftarrow \true$ \\
                            $L(u) \leftarrow L(v)$ \\
                            Add $u$ to $F'$

                    }
                } \label{algo:ldd:bfs_end}
                $F = F' \cup \{v \mid v \in B_i \mbox{ and}\,\vname{visit}[v]=\false\}$\label{algo:ldd:add}\\
                Set $\vname{visit}[v] \leftarrow \true$ for all $v \in B_i$
            }
            \Return{$L(\cdot)$}
        }
    \end{algorithm}

\subsection{Connected Components (CC)}\label{sec:cc}

Computing the connected components is one of the most widely-studied graph problems.
A recent framework \connectit{}~\cite{dhulipala2020connectit} implemented over 232 shared-memory parallel algorithms, based on
numerous previous studies both theoretically~\cite{andoni2018parallel,blelloch2012internally,gbbs2021,jain2017adaptive,ShiloachV82,shun2013ligra,SDB14,sutton2018optimizing,cao2020improved} and practically~\cite{beamer2015gap,blelloch2012internally,gbbs2021,shun2013ligra,SDB14,dhulipala2020connectit}.

Since connectivity is not the main focus of this paper, we picked one of the algorithms from \connectit{}, referred to as ``LDD-UF-JTB'', as a proof-of-concept to show the effectiveness and generality of the new techniques in this paper.
We note that none of the algorithms in \connectit{} has overwhelming advantages on all graphs. LDD-UF-JTB is one of the fastest algorithms, and our new version accelerates it by up to 3.2x compared to the original version in \connectit{}.

LDD-UF-JTB has two major components: the first step uses \defn{low-diameter decomposition} (LDD)~\cite{miller2013parallel}, and the finishing step uses the \defn{union-find structure} by Jayanti et al.~\cite{jayanti2019randomized}.
We apply our new techniques to the LDD step.
An LDD of a graph means to find a decomposition (partition of vertices) of the graph where each component has a low diameter and the number of edges crossing components is small.
This LDD step first randomly permutes all vertices, and then starts with a single source and searches out using BFS.
Then in later rounds, new sources are added to the frontier (\cref{algo:ldd:add}) in exponentially increasing batches (\cref{algo:ldd:batches}) along with the execution of BFS (\cref{algo:ldd:bfs_start} to \cref{algo:ldd:bfs_end}).
In our implementation, we increase the batch sizes by 1.2$\times$ in each round.

Our implementation replaces the BFS in \connectit{} with the more efficient reachability algorithm with VGC optimization and the parallel hash bag.
Similar to SCC, we do not need the BFS ordering in computing connectivity, so replacing BFS with (undirected) reachability searches is still correct.
In this case, our algorithm can explore more vertices in one round, which leads to fewer rounds and better parallelism.
LDD has only $O(\log n)$ rounds (\cref{line:ldd:loop}), so it is already reasonably fast.
By using local search and parallel hash bag, we further improve its performance by 1.67$\times$ (geometric mean on all graphs). We present the experimental details in~\cref{sec:exp-cc}.

\hide{
For simplicity, we will apply our new techniques to one algorithm in \connectit{} as an example both in description and in implementation. We use the algorithm in~\cite{dhulipala2020connectit}, which is referred to as ``LDD-UF-JTB'' in \connectit{}. We choose it because it has decent practical performance on most graph instances, and has theoretical guarantee in cost bounds.
We note that this algorithm is not always the fastest CC algorithm among all graphs tested in \connectit{}.
In fact, none of the algorithms in \connectit{} has overwhelming advantages, but many of them can have the best performance on certain graphs, mainly decided by certain properties of the graphs.
We note that improving CC performance on all types of graphs is not our main focus (our focus is SCC).
The discussion and experiments on CC is to show the effectiveness and generality of our proposed ideas in \cref{sec:technique}.

We will first briefly review the two key techniques that name this algorithm (LDD-UF-JTB): \defn{low-diameter decomposition} (LDD)~\cite{miller2013parallel} and a \defn{union-find structure} by Jayanti et al.~\cite{jayanti2019randomized}.
An LDD of a graph means to find a decomposition (partition of vertices) of the graph,
where (1) each component has a low diameter and (2) the number of edges cross components is small.
More formally, a $(\beta, d)$-decomposition of a graph $G=(V,E)$ is to partition of $V$ into subsets $V_1, V_2, \cdots, V_k$ such that
(1) the diameter of each $V_i$ is at most~$d$, and
(2) the number of edges $(u, v) \in E$ with endpoints in different subsets, i.e., such that $u \in V_i, v \in V_j$, and $i \neq j$, is at most $\beta m$.
We call these edges \emph{LDD cross edges} to distinguish with the cross edges in our SCC algorithm.
A parallel $(\beta, O({(\log n)}/{\beta}))$ decomposition algorithm is provided by Miller et. al.~\cite{miller2013parallel}, using $O(n+m)$ work and $O((\log^2 n)/\beta)$ span \whp.
This algorithm start with a single source and searches out using BFS.
Then in later rounds, new sources are added to the frontier (\cref{algo:ldd:add}) in exponentially increasing batches (\cref{algo:ldd:batches}) and continue BFS processes (\cref{algo:ldd:bfs_start} to \cref{algo:ldd:bfs_end}).
By controlling the speed to add new sources, the entire BFS will finish in $O((\log n)/\beta)$ rounds, leaving at most $\beta m$ edges with endpoints from different sources.

Once the LDD is computed (\cref{algo:connectivity:ldd}), the LDD-UF-JTB algorithm will examine all the LDD cross edges using a union-find structure by Jayanti et al.~\cite{jayanti2019randomized} to merge different components (\cref{algo:connectivity:union}).
The algorithm either performs finds naively, without using any path compression or uses a strategy called Find-Two-Try-Split.
Combining the two pieces together gives a practical connectivity framework.
}
\hide{
Such strategies guarantees provably-efficient bounds.
The original bound is $O(l\cdot(\alpha(n,l/(np)) + \log(np/l + 1)))$ expected work and $O(\log n)$ PRAM time for a problem instance with $l$ operations on $n$ elements on a PRAM with $p$ processors.
When translating this bound to the binary fork-join model, all $l$ operations can be in parallel in the worst case, which leads to the work bound as $O(l\log n)$.

We note that if we set $\beta=1/\log n$, the LDD takes $O(n+m)$ work and $O((\log^3 n)/\beta)$ span \whp, and the union-find part takes $O(\beta m \log n)=O(m)$ work and $O(\log^2 n)$ span.
Combining the two pieces together gives $O(n+m)$ work and $O((\log^3 n)/\beta)$ span \whp for the ``LDD-UF-JTB'' algorithm.

In our implementation, we replace the BFS in \connectit{} with the more efficient reachability algorithm with local search optimization and the hash bag data structure.
Similar to SCC, in CC we do not need the BFS ordering, but only need the (undirected) reachability information.
Therefore, using the local search does not affect the correctness of the algorithm, and allows the algorithm to explore more vertices in one round.
Our experiments show that our new implementation improves the performance of that in \connectit{} by xxx (see more details in xx).
}

\hide{
\connectit{} produces algorithms in \defn{sampling} phase and \defn{finish} phase.
In the sampling phase, it uses a subset of the edges to partition the graph into different vertex subsets, and finds the largest subset.
In the finish phase, it only runs connected components on incident edges with an endpoint not in the largest subset.
} 
\begin{algorithm}[t]
\caption{\small BGSS Algorithm for LE-Lists~\cite{blelloch2016parallelism}}
\label{algo:lelist}
\small
\SetKwFor{parForEach}{parallel\_for\_each}{do}{endfor}
\KwIn{A graph $G=(V,E)$  with $V = \{v_1,\ldots,v_n\}$}
\KwOut{The LE-lists $L(\cdot)$ of $G$}
\DontPrintSemicolon
    \smallskip
    Set $\delta(v)\leftarrow +\infty$ and $L(v) \leftarrow \emptyset$ for all $v\in V$\\
    Partition $V$ into $\log n$ batches $P_{1..\log n}$, where $|P_{i}|=2^{i-1}$\label{line:partition}\\
    \For{\upshape $i \gets 1, \dots , \log n$} {
       Apply multi-BFS from vertices in $P_i$, and let $S=\{\langle u,v,d(v,u)\rangle ~|~v\in P_i,d(v,u)<\delta(u)\}$\label{stp:dijkstra}\\
       \lparForEach {\upshape $\langle u,v,d\rangle\in S$} {
           $\delta(u)\gets \min\{\delta(u), d\}$\label{line:le-list-update-delta}
       }
       $L'(u)\gets\{(u,v,i)\in S\}$\label{line:le-list-add} \\
       Sort $L'(u)$ based on the distances in decreasing order, filter out triples that violate constraints, and append $v$ (the second element) to $L(u)$\label{line:le-list-sort}
    }

    \Return{$L(\cdot)$}
\end{algorithm}
\subsection{Algorithm on Least-Element Lists (LE-Lists)}\label{sec:lelists}

Given an undirected graph $G=(V,E)$ with $V=\{v_1, \dots, v_n\}$ in a given random total order,
a vertex $u$ is in vertex $v$'s \defn{least-element list (LE-list)} if and only if there is no earlier vertex than $u$ in $V$ that is closer to $v$~\cite{cohen1997size}.
More formally, for $d(u,v)$ being the shortest distance between $u$ and $v$, the LE-lists of $v_i$ is:
\[
	\textstyle{L(v_i)=\left\{v_j\in V ~|~ d(v_i, v_j) < \min_{1\le k < j} d(v_i, v_k)\right\}}
\]
sorted by $d(v_i, v_j)$.

LE-lists have applications in estimating the influence of vertices in a network~\cite{cohen2004,du2013,chen2009efficient}, estimating reachability set size~\cite{roditty2008improved,king2002fully},
and probabilistic tree embeddings of a graph~\cite{blelloch2017efficient,Khan2008}, which further have numerous applications.
In this paper, we focus on the unweighted LE-lists algorithm, so the distances can be computed by BFS.

The state-of-the-art parallel algorithm to compute LE-list is the BGSS algorithm (in the same paper as the BGSS SCC algorithm~\cite{blelloch2016parallelism}).
A pseudocode is given in \cref{algo:lelist}.
It first permutes the vertices $V$, divides $V$ into $\log_2 n$ batches of size 1, 2, 4, 8, ..., and processes one batch at a time.
A tentative distance $\delta(\cdot)$ is maintained for each vertex, initialized as $+\infty$.
In each batch, it runs multiple BFS from all vertices in the batch simultaneously, based on $\delta(\cdot)$ from the previous batch.
For a vertex $v\in V$, if its search reaches $u$ using a distance smaller than $\delta(u)$, the algorithm add $\langle u,v,d(v,u)\rangle$ to a set $S$.
Finally, we use $S$ to update the tentative distance $\delta(\cdot)$ in this round (\cref{line:le-list-update-delta}), and the LE-list of each vertex (\cref{line:le-list-sort,line:le-list-add}).
Blelloch et al.\ showed that running multi-BFS in $\log_2 n$ batches enables parallelism, while the work is asymptotically the same.
Each LE-list has size $O(\log n)$ \whp{}, and the entire algorithm runs $O(m\log n)$ time.
A preliminary implementation is given in ParlayLib~\cite{blelloch2020parlaylib}, using multi-BFS discussed in \cref{sec:prelim}.

We note that we can use the parallel hash bag introduced in \cref{sec:hashbag} to maintain the frontier in the multi-BFS, which avoids the second visit in multi-BFS.
VGC is not directly applicable here because we need to preserve the BFS order.
In addition, we use a parallel hash table~\cite{shun2014phase} to check if a source-target pair is already visited.
In the round $i$, if a source vertex $v$ in the current batch reaches $u$ by a distance shorter than $\delta[u]$, and if $(v, u)$ are not in the hash table, we insert $(v, u)$ to the hash table and hash bag, and pack the hash bag as the next BFS frontier.
We insert all such triples $(u,v,i)$ to a set $S$ (\cref{stp:dijkstra}), where $v$ and $u$ are described as above, and $i$ is the current round (which is also the distance between $u$ and $v$).
Our parallel LE-lists implementation outperforms the existing implementation from \parlay{} (from the authors of the BGSS LE-lists algorithm) by 4.34$\times$ on average.

\hide{
The vertices in the frontier traverse their neighbors in parallel. Suppose $u$ is a vertice in the frontier, and $v$ is its neighbor. If $\delta(u)+1 \leq \delta(v)$, then we will update $\delta(v)$ to be $\delta(u)+1$, add $(P(u), v, \delta(v))$ to the LE-List, update $P(v)$ by priority write (write\_min) $P(u)$, and add $v$ to the next frontier.

Suppose that $u$ and $v$ have the same distance to $k$. $u$ precedes $v$ in the sequential order but $v$ may visit $k$ ealier than $u$ in the parallel setting (consider $u$ are $v$ are in the same batch). In this case, we add both $(u,k, \delta(k))$ and $(v,k,\delta(k))$ to the LE-List.  After applying BFS for all the batches, we semisort the LE-List by the target, then by the distance. Then we identify elements whose target and distance are the same remove the ones with larger sources.

Notice that if $u$ and $v$ are in the same batch and $u$ precedes $v$ in the sequential order. If the distance from $u$ to $k$ is larger than that from $v$, in the standard algorithm, both $u$ and $v$ will be added to $k$'s LE-List. In our parallel algorithm, $v$ will visits $k$ earlier than $u$ in the parallel BFS order, and when $u$ visits $k$, it can not shorter the distance, so only $v$ will be added to $k$'s LE-List.  The LE-List founded by our parallel algorithm is a subset of the standard algorithm but it will not affect the applications using the LE-List.
}
 
\begin{table*}[htbp]
 \vspace{-1em}


\newcommand{\cellcolorhighlight}{\cellcolor[rgb]{ 1,  1,  0.5}}
\newcommand{\textcolorhighlight}{\textcolor[rgb]{ 1,  0,  0}}

  \centering
  \small
    \begin{tabular}{@{}l@{}l@{   }@{   }@{   }r@{   }@{   }r@{   }r@{   }@{   }r@{}r@{   }@{   }r|r@{   }r@{   }r|r@{   }r@{   }r|r@{   }r@{   }r|r}
          &\multicolumn{1}{c}{} & \multicolumn{6}{c|}{\textbf{Graph Information}} & \multicolumn{3}{c|}{\textbf{Ours}} & \multicolumn{3}{c|}{\textbf{GBBS}} & \multicolumn{3}{c|}{\textbf{Other Benchmarks}} & \multicolumn{1}{c}{\cellcolorhighlight{$\textbf{T}_\textbf{best}$}} \\

           &\multicolumn{1}{c}{} & \multicolumn{1}{c}{$\boldsymbol{n}$} & \multicolumn{1}{c}{$\boldsymbol{m}$} & \multicolumn{1}{c@{}}{$\boldsymbol{D}$} & \multicolumn{1}{c@{}}{$\boldsymbol{|\mathit{SCC}_1|}$} & \multicolumn{1}{@{  }r}{$\boldsymbol{|\mathit{SCC}_1|\%}$} & \multicolumn{1}{c|}{\textbf{\#SCC}} &\multicolumn{1}{c}{\textbf{par.}}  & \multicolumn{1}{c}{\textbf{seq.}} & \multicolumn{1}{c|}{\textbf{spd.}} & \multicolumn{1}{c}{\textbf{par.}} &\multicolumn {1}{c}{\textbf{seq.}} & \multicolumn{1}{c|}{\textbf{spd.}} & \multicolumn{1}{c}{\textbf{iSpan}} & \multicolumn{1}{c}{\textbf{MS}} &\multicolumn{1}{c|}{\textbf{SEQ}} & \multicolumn{1}{c}{\cellcolorhighlight{/ \textbf{ours}}}\\
           \midrule
    \multicolumn{1}{@{}c@{}}{\multirow{2}[0]{*}{\begin{sideways}\textbf{Social}\end{sideways}}}
           &\textbf{LJ} & 4.85M  & 69.0M     & 16     & 3,828,682 & 78.98\% & 971,232
           & \underline{0.038} & 1.06  & 27.7  & 0.118  & 1.44  & 12.1  & 0.050$^?$ & 0.141  & 2.90  & \cellcolorhighlight1.30 \\
           &\textbf{TW} & 41.7M & 1.47B & 65     & 33,479,734 & 80.38\% & 8,044,729
           & \underline{0.226}  & 14.3 & 63.2  & 0.387  & 19.7  & 50.9  & c     & 1.32  & 71.7  & \cellcolorhighlight1.71  \\
          \midrule
    \multicolumn{1}{r}{\multirow{4}[0]{*}{\begin{sideways}\textbf{Web}\end{sideways}}}
           &\textbf{SD} & 89.2M & 2.04B & 241  & 47,965,727 & 53.74\% & 39,205,039
           & 1.96  & 104 & 46.6  & 5.25  & 110   & 21.0  & 4.78$^?$ & \underline{1.86}  & 104   & \cellcolorhighlight 0.95\\
           &\textbf{CW} & 978M & 42.6B &  666   & 774,373,029 & 79.15\% & 135,223,661
           & \underline{17.6}  & 1189 & 67.4  & 40.4  & 1,166 & 28.9  & n     & n     & 589   & \cellcolorhighlight2.29\\
           &\textbf{HL14} & 1.72B & 64.4B &  793  & 320,754,363 & 18.60\% & 1,290,550,195
           & \underline{20.6}  & 1,622 & 78.8  & 67.3  & 2,041 & 30.3  & n     & n     & 620   & \cellcolorhighlight3.27\\
           &\textbf{HL12} & 3.56B & 128B  & 5,275  & 1,827,543,757 & 51.28\% & 1,279,696,892
           & \underline{95.5}   & 8528 & 89.3  & 361   & 7,022 & 19.5  & n     & n     & 1822  & \cellcolorhighlight 3.78\\
          \midrule
    \multicolumn{1}{r}{\multirow{8}[0]{*}{\begin{sideways}\textbf{KNN}\end{sideways}}}
           &\textbf{HH5} & 2.05M & 10.2M & 980  & 257,914 & 12.59\% & 94,010
           & \underline{0.208} & 3.10  & 14.9  & \textcolorhighlight{3.95}  & 3.51  & 0.89   & \textcolorhighlight{0.791}  & \textcolorhighlight{2.21}     & 0.449  & \cellcolorhighlight 2.16 \\
           &\textbf{CH5} & 4.21M & 21.0M &  4,550  & 497,331 & 11.82\% & 248,227
           & \textcolorhighlight{0.557}  & 5.83  & 10.5   & \textcolorhighlight{8.39}  & 5.84  & 0.70   & \textcolorhighlight{2.15}  & \textcolorhighlight{17.6}  & \underline{0.427}  & \cellcolorhighlight 0.77  \\
           &\textbf{GL2} & 24.9M & 49.8M & 4,142  & 5,368 & 0.02\% & 9,705,931
           & \underline{0.598}  & 39.1  & 65.3  & 3.00  & 82.4  & 27.5  & \textcolorhighlight{t}     & \textcolorhighlight{8.36}     & 3.39  & \cellcolorhighlight 5.01 \\
           &\textbf{GL5} & 24.9M & 124M  & 12,059 & 860,403 & 3.46\% & 3,198,626
           & \underline{0.865}  & 45.8  & 53.0  & \textcolorhighlight{10.5}  & 91.0  & 8.69   & \textcolorhighlight{t}     & \textcolorhighlight{19.1}     & 4.83  & \cellcolorhighlight 5.58\\
           &\textbf{GL10} & 24.9M & 249M  & 4,531  & 3,042,330 & 12.23\% & 326,811
           & \underline{1.49}  & 61.6  & 41.4  & \textcolorhighlight{12.3}  & 76.5  & 6.24   & \textcolorhighlight{35.2}  & 7.14     & 9.30  & \cellcolorhighlight 4.79\\
           &\textbf{GL15} & 24.9M & 373M  & 5,491  & 3,239,156 & 13.02\% & 187,646
           & \underline{2.09}  & 75.5  & 36.1  & \textcolorhighlight{13.7}  & 84.5  & 6.15   & \textcolorhighlight{29.4}  & 10.6     & 11.3  & \cellcolorhighlight 5.06 \\
           &\textbf{GL20} & 24.9M & 498M & 5,275 & 3,336,963 & 13.41\% & 128,021
           & \underline{2.38}  & 86.0  & 36.1  & \textcolorhighlight{14.5}  & 96.6  & 6.68   & \textcolorhighlight{27.3}  & 12.3     & 13.3  & \cellcolorhighlight 5.18 \\
           &\textbf{COS5} & 321M & 1.61B & 1,148 &301,413,787 & 93.88\% & 2,273,690
           & \underline{3.22}  & 284   & 88.2  & 12.0  & 367   & 30.7  & \textcolorhighlight{t}     & 57.4     & 189   & \cellcolorhighlight 3.72\\
          \midrule
    \multicolumn{1}{@{}c@{}}{\multirow{4}[1]{*}{\begin{sideways}\textbf{Lattice}\end{sideways}}}
           &\textbf{SQR} & 100M & 300M & 10,002  & 99,101,606 & 99.10\% & 829,495
           & \underline{0.577}  & 24.7  & 42.8  & 11.1  & 28.5  & 2.57   & 4.45$^?$ & 12.6  & 15.5  & \cellcolorhighlight 7.72 \\
           &\textbf{REC} & 10M & 30M  & 5,946  & 9,890,647 & 98.91\% & 101,059
           & \underline{0.117}  & 2.08  & 17.8  & \textcolorhighlight{3.82}  & 2.14  & 0.56   & 1.19$^?$ & \textcolorhighlight{5.24}  & 1.57  & \cellcolorhighlight 10.2 \\
           &\textbf{SQR'} & 100M & 120M & 51   & 58    & 0.00\% & 78,052,793
           & \underline{1.38}  & 105   & 76.3  & 4.76  & 243   & 51.0  & \textcolorhighlight{26.4$^?$} & 3.19     & 6.90  & \cellcolorhighlight 2.31\\
           &\textbf{REC'} & 10M & 12M & 80   & 42    & 0.00\% & 7,819,050
           & \underline{0.159}  & 9.38  & 59.0  & \textcolorhighlight{1.00}  & 18.7  & 18.8  & \textcolorhighlight{0.851$^?$} & \textcolorhighlight{0.645}     & 0.60  & \cellcolorhighlight 3.75\\
    \bottomrule
    \end{tabular}
    \vspace{-1.5em}
    \caption{\small \label{tab:graphfull} \textbf{The running times (in seconds) of all tested algorithms on SCC.} $n=$ number of vertices. $m=$ number of edges. $D=$ estimated diameter (a lower bound of the actual value). $|\mathit{SCC}_1|=$ largest strongly connected component (SCC) size. $|\mathit{SCC}_1|\%= |\mathit{SCC}_1|/n$, ratio of the largest SCC. \#SCC $=$ number of SCCs.
    ``iSpan'' $=$ iSpan algorithm~\cite{ji2018ispan}. ``MS'' $=$ Multi-step algorithm~\cite{slota2014bfs}. ``SEQ'' $=$ classic sequential SCC~\cite{tarjan1972depth}.
    ``par.'' $=$ parallel running time (on 192 hyperthreads). ``seq.''$=$ sequential running time. ``spd.''$=$ self-relative speedup.
    ``?''$=$ number of SCCs is different from other implementations.
    \hide{
      ``!''$=$ largest SCC size is different from other implementations.}
    ``t''$=$ timeout (more than 5 hours). ``c''$=$ crash. ``n''$=$ no support.
    We set $\tau=2^9$.
    The fastest runtime for each graph is underlined.
    Red numbers are parallel running times \emph{slower than \seq{}}.
    }
    \vspace{-1em}
\end{table*}%

\section{Experiments}\label{sec:exp}

\myparagraph{Setup.}
We run our experiments on a 96-core (192 hyperthreads) machine with four Intel Xeon Gold 6252 CPUs and 1.5 TB of main memory.
We implemented all algorithms in C++ using \parlay{}~\cite{blelloch2020parlaylib} for fork-join parallelism and parallel primitives (e.g., sorting).
We use \texttt{numactl -i all} for parallel tests to interleave the memory pages across CPUs in a round-robin fashion.
All reported numbers are the average running time of the last five out of six runs.


We use $\tau=2^9$ in all tests except for those in \cref{fig:tau} which studies the choice of $\tau$.
We tested 18 directed graphs, including social networks, web graphs, \knn{} graphs, and lattice graphs.
All social, web, and \knn{} graphs are real-world graphs, with up to 3.6 billion vertices and up to 128 billion edges.
The lattice graphs are generated by a similar model in \cite{de2018percolation},
which uses SCC to study the percolation on isotropically directed lattices.
Basic information on the graphs is given in \cref{tab:graphfull}.
For social graphs, we use \emph{LiveJournal} (LJ)~\cite{backstrom2006group} and \emph{Twitter} (TW)~\cite{kwak2010twitter}.
For web graphs~\cite{webgraph}, we use \emph{sd-arc} (SD), \emph{ClueWeb} (CW), \emph{Hyperlink12} (HL12) and \emph{Hyperlink14} (HL14).
\knn{} graphs are widely used in machine learning algorithms~\cite{Maier2009,Franti2006,Lucinska2012,karypis1999chameleon,Tenenbaum2000,Hautamaki2004,Paredes2005,Chavez2010,Sebastian2002}.
In \knn{} graphs, each vertex is a multi-dimensional data point and has $k$ edges pointing to its $k$-nearest neighbors (excluding itself).
We use \emph{Household} with $k=5$ (HH5)~\cite{uciml,wang2021geograph}, \emph{Chemical} with $k=5$ (CH5)~\cite{fonollosa2015reservoir,wang2021geograph}, \emph{GeoLife} with $k=2,5,10,15,20$ (GL2, GL5, GL10, GL15, GL20)~\cite{geolife,wang2021geograph}, and \emph{Cosmo50} with $k=5$ (COS5)~\cite{cosmo50,wang2021geograph}.
We also created four lattice graphs~\cite{de2018percolation},
including two $10^4 \times 10^4$ 2D-lattices (SQR and SQR'),
and two $10^3 \times 10^4$ 2D-lattices (REC and REC').
Each row and column in the lattice graphs are circular.
In SQR and REC, for each vertex $u$ and its adjacent vertex $v$, we add a directed edge from $u$ to $v$ with probability 0.5,
and from $v$ to $u$ otherwise, then remove duplicate edges.
In SQR' and REC', for each vertex $u$ and each of its adjacent vertex $v$, we create an edge from $u$ to $v$ with probability 0.3, and from $v$ to $u$ with probability 0.3, and create no edge with probability 0.4, then remove duplicate edges.

To test our connectivity and LE-lists algorithms, we symmetrize all 18 directed graphs and use 4 more real-world undirected graphs, com-orkut (OK)~\cite{yang2015defining}, Friendster (FT)~\cite{yang2015defining}, RoadUSA (USA)~\cite{roadgraph}, and Germany (GE)~\cite{roadgraph}.
Graph details are given in \cref{tab:cclelists}.

We call the social and web graphs \emph{low-diameter graphs} as they usually have low diameters (roughly polylogarithmic in $n$). We call the \knn{} and lattice graphs \emph{large-diameter graphs} as their diameters are large (roughly $\Theta(\sqrt{n})$). When comparing the \emph{average} running times across multiple graphs, we always use the \defn{geometric mean}.

\myparagraph{Baseline Algorithms.} We call all existing algorithms that we compare to the \defn{baselines}.
We compare the number of SCCs and the largest SCC size reported by each algorithm with \seq{} to verify correctness.
For SCC, we compare to \gbbs{}~\cite{gbbs2021, dhulipala2020graph}, \ispan{}~\cite{ji2018ispan}, and \multistep{}~\cite{slota2014bfs}.
\gbbs{} also implements the BGSS algorithm, so we also compare our breakdown and sequential running times with \gbbs{}.
We also implemented and compared to Tarjan's sequential SCC algorithm~\cite{tarjan1972depth}.
We call it \seq{}.
On six graphs, \ispan{}'s results are off by 1, noted with ``?'' in \cref{tab:graphfull}.
\hide{
  For all the eleven graphs that \ispan{}'s can run out, they are either has wrong number of SCCs, noted as ``?'';
  or has wrong lagest SCC size, noted with ``!'' in \cref{tab:graphfull}.
}
(We communicated with the authors but could not correct it.)
\multistep{} and \ispan{} do not support CW, HL12, and HL14 because they have more than $2^{32}$ edges.
For connectivity, we apply our two techniques to the LDD-UF-JTB algorithm in \connectit{}~\cite{dhulipala2020connectit, dhulipala2020connectit} and compare it to the original implementation
in \connectit{}. For \lelist{}s, we compare to \parlay{}, which is
the only public parallel LE-lists implementation to the best of our knowledge. 





We first summarize the overall performance of the algorithms and scalability tests in \cref{sec:exp:overall}.
Next, we show some experimental studies on performance breakdown in \cref{sec:exp:breakdown}, and an in-depth study of \VGC{} in \cref{sec:exp:tech}.
Finally, we provide a brief summary of our experimental results for connectivity and LE-lists in \cref{sec:exp-cc-lelists}.

\hide{
\begin{figure}
  \centering
  \vspace{-2em}
  \includegraphics[width=\columnwidth]{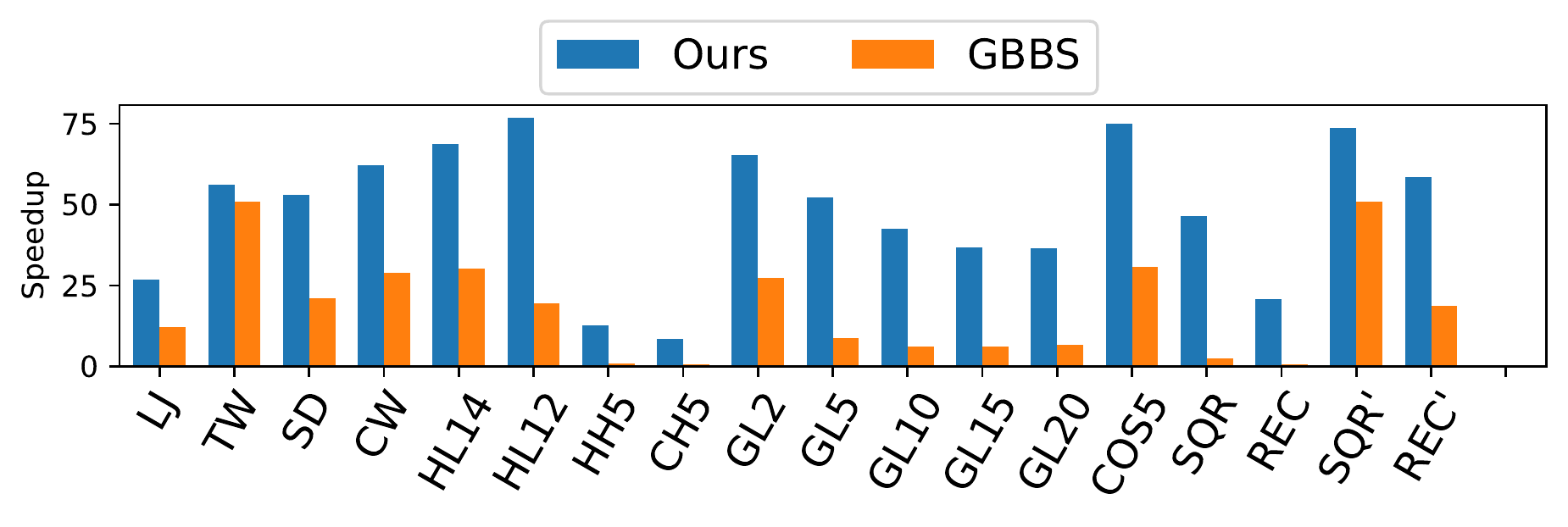}
  \captionof{figure}{\small \textbf{Self-relative speedup for our algorithms and the algorithms in GBBS.}Much of our performance improvement comes from better parallelism. \label{fig:speedup}}
\end{figure}
}

\subsection{Overall Performance}
\label{sec:exp:overall}

We show the running times in \cref{tab:graphfull} and a heatmap in \cref{fig:heatmap}.
We mark the parallel running times \emph{slower than the sequential algorithms} in red in \cref{tab:graphfull}.
Our implementation is almost always the fastest except on SD and CH5.
On CH5, we are 23\% slower than \seq{}.
CH5 has a very large diameter (4000+) compared to its small size (4M vertices), and none of the parallel implementations outperform \seq{}. 
On SD, we are only 4\% slower than \multistep{} with $\tau=2^9$. SD is one of the graphs that are dense and potentially has good parallelism, and thus may prefer smaller $\tau$.
As we will show in \cref{fig:tau}, using $\tau\le 2^8$ will achieve a better performance than all existing implementations on SD, but we keep the results in \cref{tab:graphfull} all using $\tau=512$ for simplicity.
The highlighted columns in \cref{tab:graphfull} show the speedup of our algorithms to the \emph{best} baseline (including \seq{}) on each graph.
Compared to the best baseline, we are up to 10.2$\times$ faster and 3.1$\times$ faster on average.
All the implementations perform favorably on all low-diameter graphs (5--317$\times$ faster than \seq{}).
Conceptually, all parallel implementations first use BFS-like algorithms to find the largest SCC.
On all but one low-diameter graph, the largest SCC contains more than 50\% vertices.
Therefore, using a parallel BFS (with optimizations such as dense modes) gives decent performance.
Even so, using \hashbag{s} and \VGC{} still gives good performance on low-diameter graphs, and
we are faster (up to 3.8$\times$) than the best baseline on all but one graphs.
\revision{
One interesting finding is that on TW, our implementation, \gbbs{}, and \multistep{}
are faster than \seq{} even running sequentially.
Similar trends (running the parallel algorithm sequentially is faster than the classic sequential algorithm) have been observed in other BFS-like graph algorithms~\cite{shun2013ligra}.
This is mostly due to the dense-mode optimization as described in \cref{sec:implementation}.2.
When the frontier size is large, triggering the dense mode can skip many edges, so the number of visited edges can be fewer than $\Theta(m)$ as in the standard sequential solution.
Another reason is that our implementation (and \gbbs's) using BFS
is more I/O-friendly than Tarjan's DFS-based algorithm.}
\hide{
  In dense mode, every vertex in $V$ that has not been visited explores its in-coming neighbors. Once they find a neighbor in the current frontier, it will mark itself as visited, put itself to the next frontier, and stop exploring its remaining neighbors.
}

Our algorithm has dominating advantages on large-diameter graphs.
On \knn{} and lattice graphs, existing parallel implementations are slower than the sequential algorithms in 24 out of 36 tests.
If we take the average time of the baseline parallel algorithms for \knn{} and lattice graphs,
all of them are slower than \seq{} (see the ``MEAN'' columns in \cref{fig:heatmap}).
In comparison, our implementation is 5.3$\times$ better than \seq{} on \knn{} graphs and 9.1$\times$  better on lattice graphs.
We believe the high performance is from good parallelism.
We study the scalability of our algorithm in the next paragraph.

\begin{figure}
  \centering
    \includegraphics[width=\columnwidth]{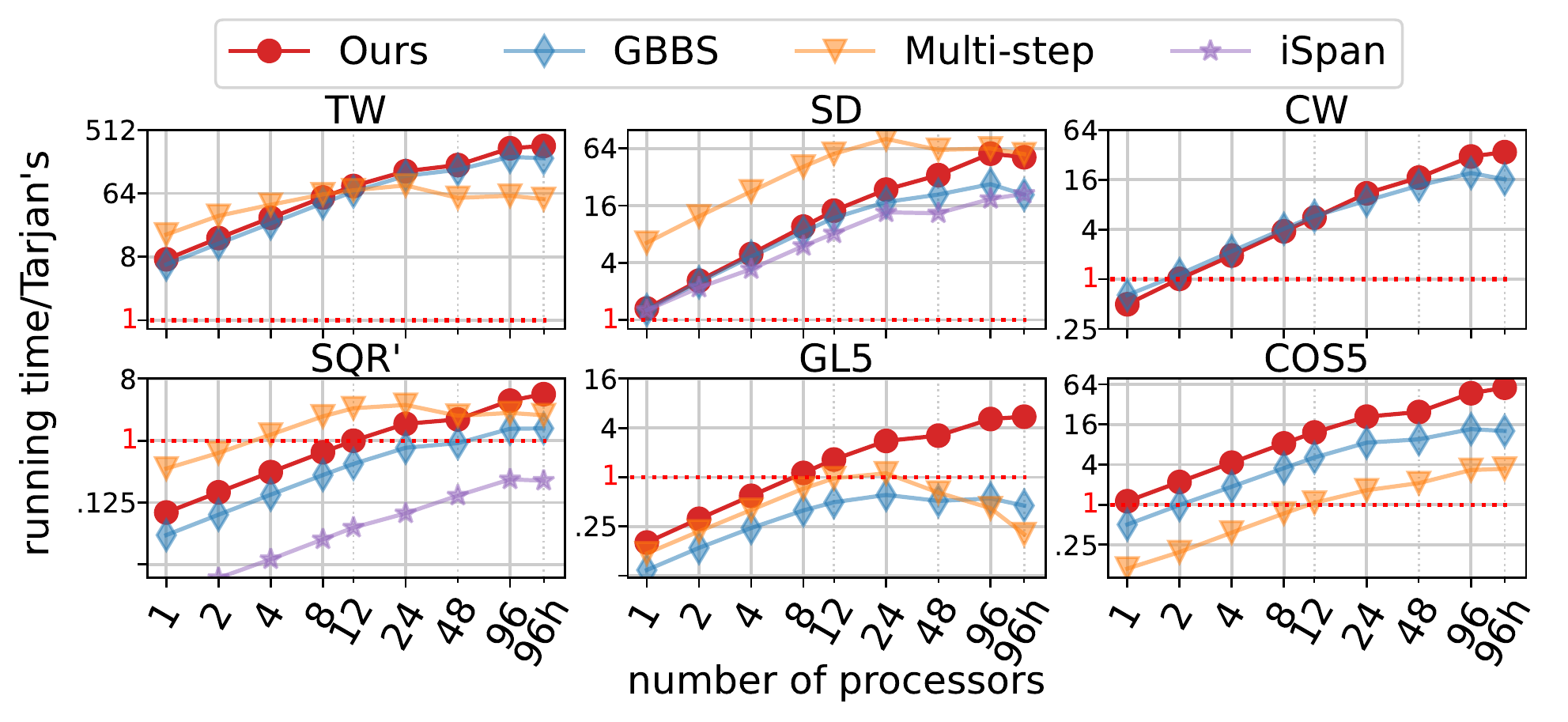}
  \caption{\small\revision{ \textbf{Comparing speedup over Tarjan's sequential algorithm on different algorithms on different number of processors.}} Larger is better. The red horizontal line indicates the running time of Tarjan's algorithm. We omit an algorithm on a graph in cases of crash/timeout/no support. }
  \label{fig:scale_all}
  \vspace{-.1in}
    \begin{minipage}[h]{.62\columnwidth}
      \includegraphics[width=\columnwidth]{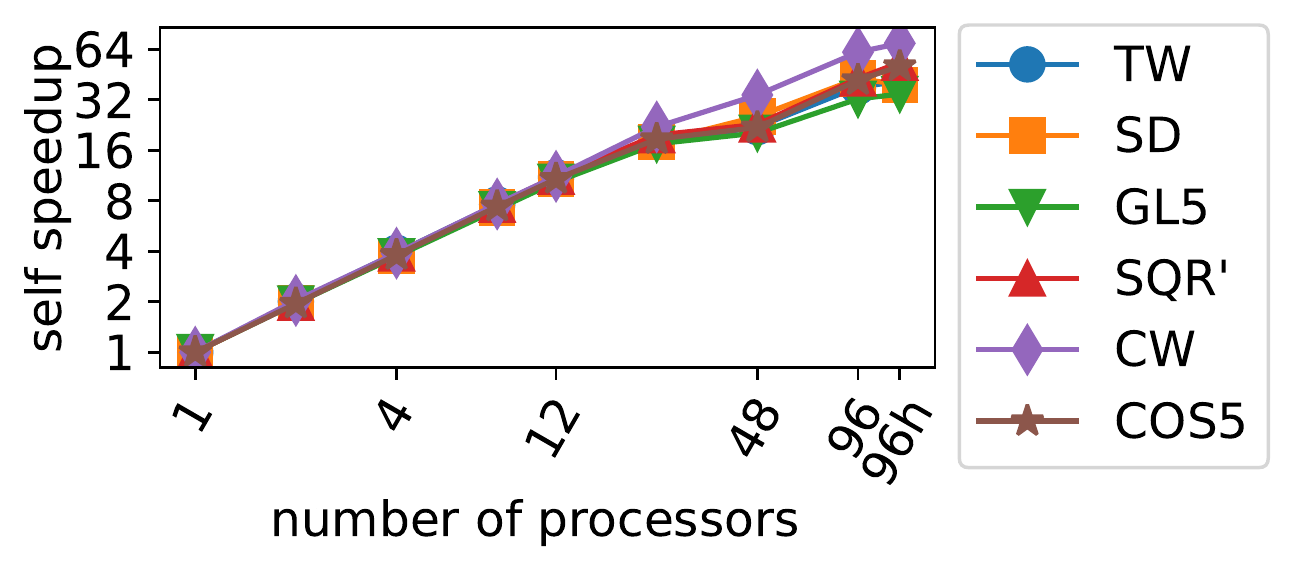}
    \end{minipage}\hfill
    \begin{minipage}[h]{.36\columnwidth}
    \caption{\small \textbf{Self-relative speedup on different processor counts, with $\tau=2^9$ on six graphs.}
    \label{fig:scale}
    }
    \end{minipage}
\end{figure}


\myparagraph{Scalability Tests.}
\revision{
  We show the speedup of four algorithms (ours, \gbbs{}, \ispan{}, \multistep{}) over the sequential Tajan's algorithm on six representative graphs in \cref{fig:scale_all}. We vary the number of processors from 1 to 96h (192 hyperthreads). The red horizontal dot lines represent the running time of Tarjan's algorithm (\seq{}), above which means faster than Tarjan's algorithm.

  On low-diameter graphs (TW, SD, and CW), all algorithms show reasonably good speedup.
  On large-diameter graphs (SQR', GL5, and COS5), our algorithm achieves significantly better scalability than the baselines.
  Our algorithm is the only one that achieves almost linear speedup on all the six graphs.
  For all the other algorithms, their performance stops increasing (dropped or flattened) with more than $24$ threads on one or more graphs.
  \multistep{} shows good performance on SD, and has better performance than our algorithm especially on a small number of threads.
  However, \multistep{} does not scale well to more processors on most of the graphs.



}

We also show the self-speedup of our algorithms on six graphs in \cref{fig:scale}. We vary the number of processors from 1 to 96h (192 hyperthreads). Due to page limitation, we do not show the curves for all graphs, but the self-speedup on all graphs on 96h (192 hyperthreads) is given in \cref{tab:graphfull}.
Our self-speedup is more than 35 except for some very small graphs.
This indicates that high parallelism is a crucial factor contributing to the high performance of our code.
Compared to \gbbs{}, the fastest previous parallel SCC implementation, our self-speedup is 1.2--32$\times$  better.
With limited parallelism, \gbbs{} can be slower than \seq{} on 8 out of 14 large-diameter graphs---the BGSS SCC algorithm has $O(m\log n)$ work compared to $O(m)$ of Tarjan's sequential algorithm, so with poor self-speedup, the parallelism cannot make up the factor of $O(\log n)$ loss in the total work.

We believe that our good performance comes from using \hashbag{s} (saving work on processing sparse frontiers)
and \VGC{} (reducing the number of rounds in reachability searches and improving parallelism).
We will discuss more details by comparing the performance breakdown with \gbbs{} in \cref{sec:exp:breakdown}, and studying the benefit brought up by \VGC{} in \cref{sec:exp:tech}.

\subsection{Performance Breakdown}
\label{sec:exp:breakdown}
\begin{figure*}[th]
  \centering
  \vspace{-1em}
  \includegraphics[width=\textwidth]{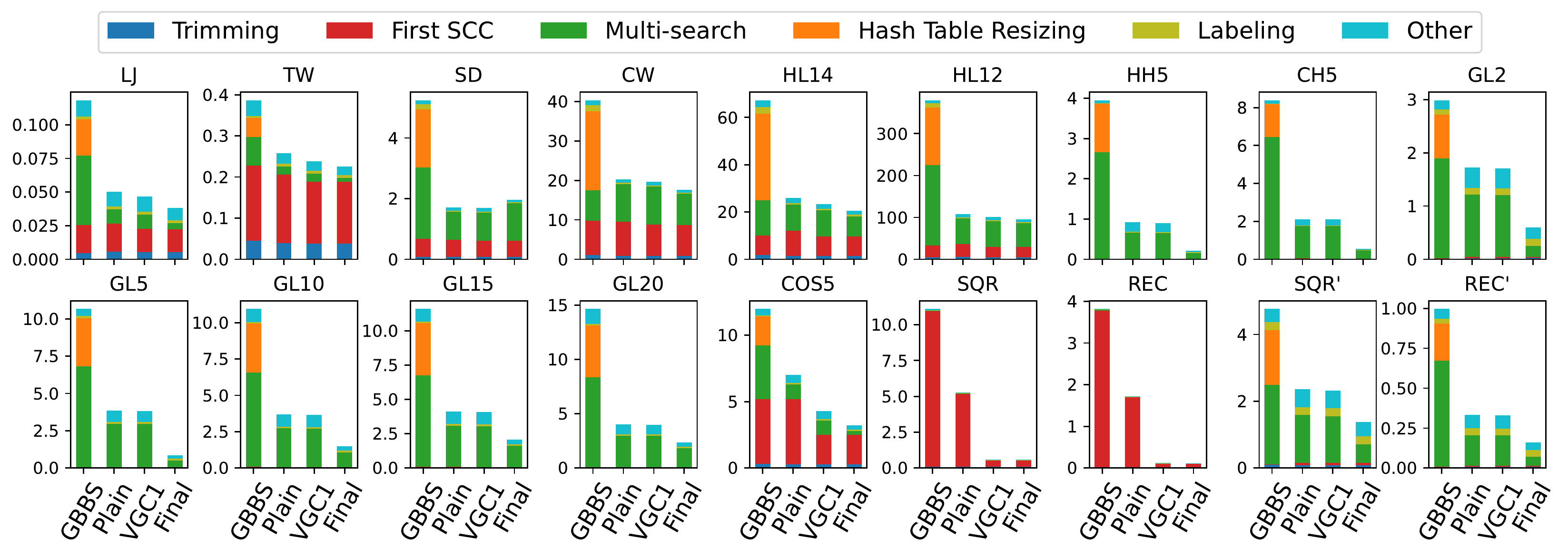}
  \caption{ \revision{\small \textbf{SCC breakdown time (in seconds).}}  $y$-axis is the running time in seconds.  All settings are the same as \cref{tab:graphfull}.
  ``Plain''$=$ our implementation with hash bags but not local search. ``VGC1''$=$ adding local search to the single-reachability search. ``Final''$=$ our final implementation with local search enabled on both single- and multi-reachability searches. The numbers on the top show the speedup of our implementations over GBBS (the first bar).
  \label{fig:SCC_Total_Break}}
  \vspace{-.1in}
    \includegraphics[width=\textwidth]{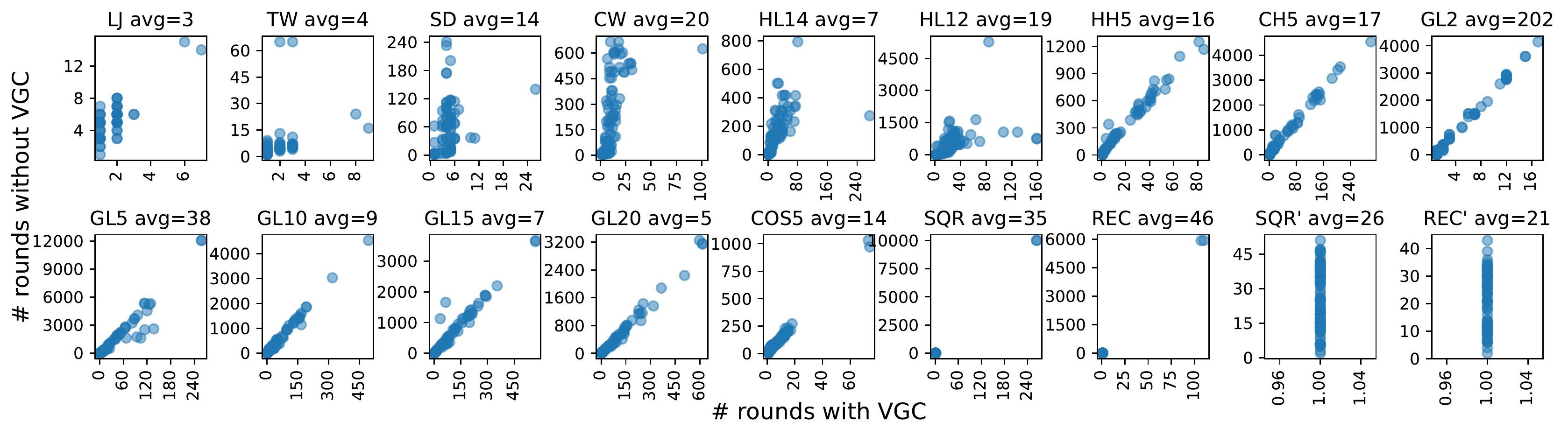}
  \caption{\small \textbf{Number of rounds with and without local search for each batch.} All settings are the same as \cref{tab:graphfull}.
  Each data point $(x,y)$ means that in one reachability search, $y$ rounds are needed using local search, and $x$ rounds are needed without local search.
  The number ``avg'' for each graph is the average number of $k=y/x$ for all data points, which means, on average, using local search reduces the number of rounds needed by a factor of $k$.
  \label{fig:SCC-rounds} }
  \vspace{-.25in}
\end{figure*}

To better understand the performance of our algorithm, we compare the performance breakdown with \gbbs{} in \cref{fig:SCC_Total_Break} since
\gbbs{} is also based on the BGSS algorithm and we have similar framework.
We compare the running time in six parts (see \cref{sec:implementation}):
1) \defn{Trimming}: trimming vertices with no in- or out-degrees; 2) \defn{First SCC}: finding the first SCC using two single-reachability searches; 3) \defn{Multi-search}: all multi-reachability searches;
4) \defn{Hash Table Resizing}: resizing the hash table storing reachability pairs; 5) \defn{Labeling and Others}: assigning labels to vertices and other costs.
We show the breakdown figure for all graphs in \cref{fig:SCC_Total_Break}.
We tested three versions of our algorithm: the \defn{plain} version uses parallel \hashbag{s} without \VGC{},
the ``\defn{VGC1}'' version enables \VGC{} in single-reachability to find the first SCC,
and the ``\defn{final}'' version fully enables \VGC{} in both single- and multi-reachability search.
\revision{We note that some graphs requires more time on \emph{First-SCC} while the others spent more time on \emph{Multi-search}
because of different graph patterns, which is indicated by the value of $|SCC_1|\%$ as shown in \cref{tab:graphfull}.  }

One straightforward improvement of our algorithm is from our better heuristic to estimate the hash table size (see details in \cref{sec:implementation}), which avoids frequent size predicting and hash table resizing.
This can be seen by comparing the time of ``hash table resizing'' (green bars) for \gbbs{} and our versions.
This optimization saves much time on almost all graphs.
In the following, we use the breakdown results to illustrate the performance improvement from our two main techniques: the \hashbag{}
and VGC.

\hide{Even our \emph{plain} version is better than \gbbs{} on almost all graphs for two reasons.
First, using parallel \hashbag{s} saves time in maintaining the frontiers by avoiding the edge-revisiting scheme.
\gbbs{} has to visit the frontiers' edges twice to compute the next frontier, but such redundant work is avoided using parallel hash bags.
This can be seen by comparing the time of ``first-SCC'' (red bars) on SQR and REC, and ``multi-search'' (green bars) for TW, GL10, SQR', etc.
Another saving is from our better heuristic to estimate the hash table size (see details in \cref{sec:implementation}), which avoids frequent hash table resizing. This can be seen by comparing the time of ``hash table resizing'' (orange bars) for \gbbs{} and our \emph{plain} version.
This optimization saves much time on almost all graphs.
}
\revision{
  \myparagraph{Evaluating \hashbag{s}.} 
  Parallel \hashbag{s} improve the performance by maintaining the frontiers without the edge-revisiting scheme.
  Note that both our algorithm and \gbbs{} use the BGSS algorithm and perform the same computation in each round, but \gbbs{} uses edge-revisiting and our algorithm avoids that by using the \hashbag{}.
  Therefore, we compare our \emph{plain} version (i.e., disabling VGC) with \gbbs{} to evaluate the improvement from \hashbag{s},
  because the major difference between them is the use of \hashbag{s}.
  We also exclude the hash table resizing time (the green bars) for fair comparison.
  On all but one graphs, using \hashbag{s} greatly improve the performance in single- and/or multi-reachabilty searches.
  Comparing the total running time of reachability searches (red and blue bars), our algorithm is up to 4$\times$ faster than \gbbs{} (2$\times$ faster on average),
  and the major improvement is from \hashbag{s}.
}

\revision{\myparagraph{Evaluating VGC.}} On top of our \emph{plain} version, \VGC{} improves the performance on almost all graphs.
Note that for low-diameter graphs, since the number of needed rounds is small, there is sufficient parallelism to explore.
Therefore, using \VGC{} does not improve the performance too much.
As mentioned, on SD, the performance drops slightly using local search with $\tau=2^9$, but
using smaller values of $\tau$ can still improve the performance (see \cref{sec:exp:tech}).
To keep the parameter setting simple, we still report the numbers with $\tau=2^9$ in \cref{tab:graphfull,fig:SCC_Total_Break}.
The large-diameter graphs with the largest SCC as 50\% of the graph (e.g., COS5, REC, and SQR) greatly benefit from \textit{VGC1} (using \VGC{} in the single-reachability search to find the first SCC). Comparing ``first-SCC'' of \emph{plain} and \emph{VGC1}, \VGC{} makes the performance 2.2--17$\times$ faster in the single-reachability search on COS5, SQR, and REC.
All the other large-diameter graphs get significant improvement from \textit{VGC1} to \textit{final} (using \VGC{} also in multi-reachability searches).
For all large-diameter graphs, the ``multi-search'' time in \emph{final} is smaller than that in \emph{VGC1} (1.43--14.7$\times$ improvement).
As we will show in \cref{sec:exp:tech}, this is because \VGC{} reduces the number of rounds in reachability searches by 3--200$\times$.

In summary, comparing our \emph{plain} version with \gbbs{}, we can see that \hashbag{} and our heuristic on hash table resizing improves the performance over \gbbs{} by about 1.5--4.3$\times$.
Comparing \emph{plain} with \emph{VGC1} and \emph{final}, we can see that VGC improves the performance in both single- and multi-reachability queries by up to 14.7$\times$.

\subsection{In-depth Performance Study of \VGC{}}
\label{sec:exp:tech}

\myparagraph{Reduced Number of Rounds.}
We study the improvement of \VGC{}
by reporting the number of rounds in the reachability searches with or without \VGC{} (see \cref{fig:SCC-rounds}).
In a given graph, for all single- and multi-reachability searches in the SCC algorithm, we record the number of rounds $y$ needed in plain BFS
and the number of rounds $x$ with \VGC{} enabled.
We then plot all such points $(x,y)$ on a 2D plane to illustrate the effectiveness of local search, shown in \cref{fig:SCC-rounds}.
We also report the average ratio of $y/x$ on the top of each figure.
The conceptual ``slope'' indicated by the points illustrates the factor in the reduction of the number of rounds by using local search.
For most of the graphs, especially the \knn{} graphs, thousands of rounds were needed in each multi-reachability search using BFS.
With \VGC{}, the number of rounds is mostly within 100.
Even for the cases where BFS only needs a few (10--100) rounds, \VGC{} still reduces the number of rounds
to be within 10 rounds (e.g., LJ, TW, COS5, SQR', REC').
On all graphs, the number of rounds is reduced by 3--200$\times$. As a result, the scheduling and synchronization overhead is greatly reduced.

\hide{
On SQR, the big improvement is in the first single-source search, where the number of rounds reduces from $10^4$ to about 100.
For the multi-searches, the number of rounds always reduces to one since BFS also just needs a few rounds.
This is consistent with the result in \cref{fig:SCC_breakdown}, where GL5 benefit mostly from the local search on multi-reachability queries, and SQR benefit mostly from adding local search to the first round.
}

\hide{
\myparagraph{Hash Bag Study.} To understand performance gain from hash bags, we compare the sequential running time of our SCC implementation with \gbbs{}. This is because we use a similar high-level framework as \gbbs{}, and the sequential running time reflects the work of each algorithm. On all large-diameter graphs (where dense mode is hardly used), our sequential time is much faster than theirs (up to more than 2x). We believe this is mostly due to the use of hash bags---we do not need to process the frontier multiple times.
We note that this is not a direct measurement as it is also affected by other techniques such as local search, but conceptually local search should not make much difference in work.
}

\begin{figure}
  \centering
  \vspace{-.5em}
  \includegraphics[width=1\columnwidth]{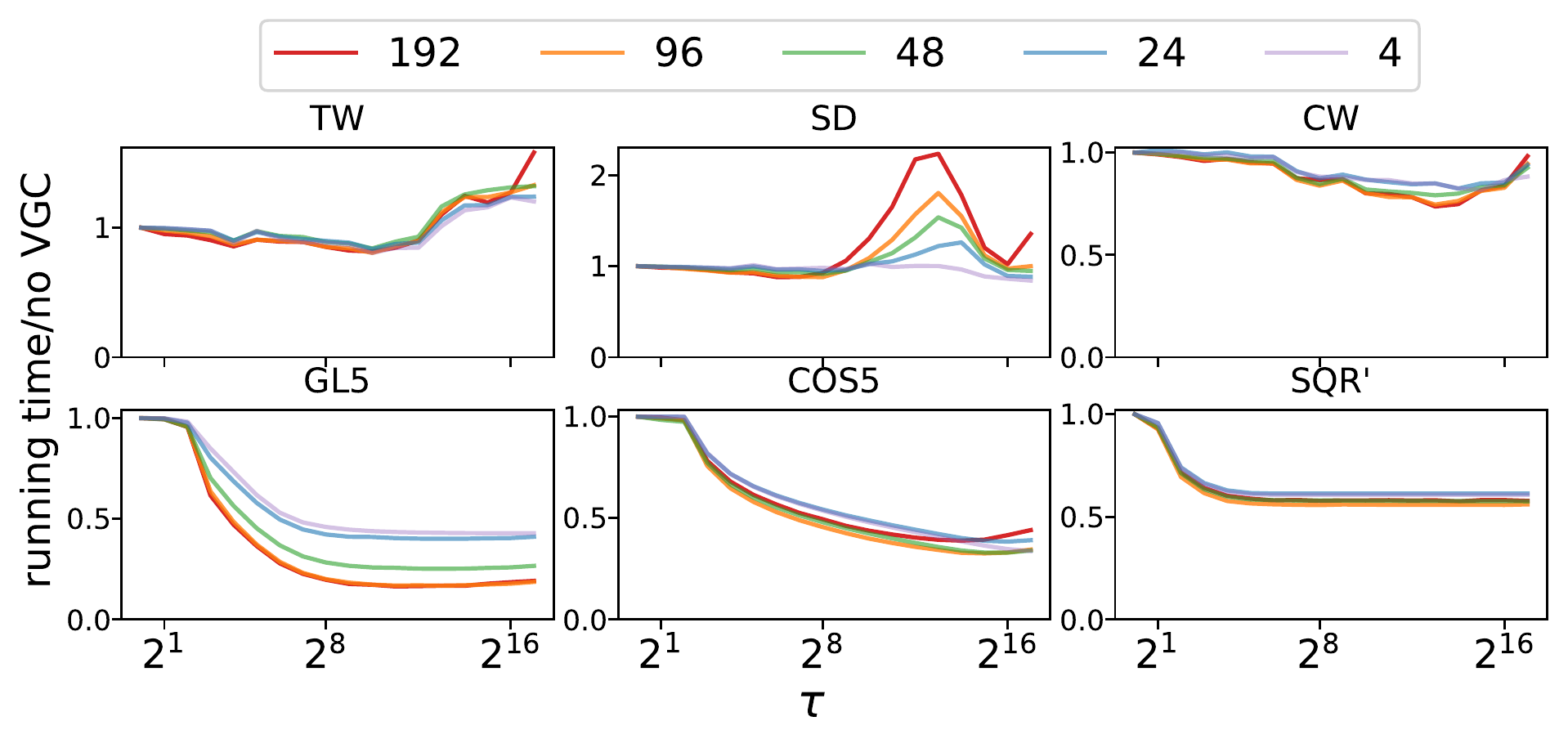}
  \vspace{-2em}
  \caption{\small \revision{\small \textbf{Relative running time to $\tau = 1$ on six graphs with $\tau$ range from $2^{0}$ to $2^{17}$, $4$ to $192$ hyperthreads (96h).}} LJ has similar trends as TW. HL12 and HL14 show similar trends as CW.
  All \knn{} and lattice graphs show similar trends as GL5, COS5 and SQR'.}
  \label{fig:tau}
  \vspace{-.15in}
\end{figure} 
\myparagraph{Choice of Parameter $\tau$.} To understand the impact of $\tau$ values on performance, \revision{we record the speedup over our \textit{plain} version (i.e., no VGC) with different values of $\tau=1$ to $2^{17}$, and under different number of processors from 96h (192 hyperthreads) to $4$}. For page limitation, we show six graphs in \cref{fig:tau} (at least one in each graph type). All the other graphs showed similar trends as one of the six examples.
\revision{We start from the curves of 192 hyperthreads on different graphs.}
On all graphs except for LJ, TW and SD, we observe improvement as long as \VGC{} is used (compared to plain BFS where $\tau=1$) for any $1<\tau \le 2^{16}$.
Overall, the performance is not sensitive (and is always better than $\tau=1$) in a large parameter space $2^6\le \tau\le 2^{12}$ on almost all graphs.
Based on the results, we set $\tau=2^9$ as it gives the best overall performance across all graphs.
Using $\tau=2^9$, SD is the only graph that has worse performance than $\tau=1$. We note that SD still benefits from \VGC{} with $\tau\le 2^8$.
Note that using larger $\tau$ suppresses parallelism, and for dense graphs with sufficient parallelism, a smaller $\tau$ may perform better.
\revision{
  Although we choose the best parameter by using experiments on 96h, we also test how different numbers of processors $P$ affect the choices of $\tau$.
  Interestingly, the trends are usually similar regardless of the number of threads used.
  With a smaller value of $P$, the performance is less sensitive to the $\tau$ value.
  This is because $\tau$ trades off between scheduling overhead and load balancing,
  and both affect the performance more when $P$ is large.}

We believe that an interesting future work is to set $\tau$ dynamically to achieve the best benefit from \VGC{}, possibly based on the sparseness of the graph and the potential parallelism, e.g.,
the edge-vertex ratio $m/n$, the number of processors $P$, or the frontier size.

\hide{
\subsection{Experiments on Connectivity and LE-Lists}\label{sec:exp-cc-lelists}

The running time of Connectivity and \lelist{s} are shown in (\cref{tab:cclelists}) and full discussions for connectivity and \lelist{s} are shown below.

Our CC algorithm is based on the LDD-Union-Find algorithm in \connectit{}~\cite{dhulipala2020connectit} (integrated in \gbbs{}),
using hash bags and the local search optimization.
On low-diameter scale-free networks, the most costly rounds are in the dense mode, so our CC has very similar performance as \gbbs{}.
On large-diameter graphs, our CC is 1.98$\times$ faster than GBBS on average, mainly due to our local search optimization.
Our CC is faster than GBBS on 20 out of 24 graphs.

For \lelist{s}, we implemented the parallel BGSS \lelist{s} algorithm~\cite{blelloch2016parallelism}.
Although \lelist{s} have numerous applications, we note that there was no previous parallel implementation,
mainly due to the lack of a suitable bag-like data structure.
Using our parallel hash bag in~\cref{sec:hashbag}, our parallel \lelist{s} implementation can achieve 62$\times$ speedup on average on scale-free graphs compared to the sequential Cohen's algorithm~\cite{cohen1997size}.
The speedup for large-diameter is lower (15.4x), since local search optimizations are not applied here.
Our implementation is the first parallel implementation for \lelist{s}.
}

\newcommand{\tableside}{1mm}
\begin{table}\centering
\small
\begin{tabular}{@{}r@{ }@{ }r|r@{  }r@{  }@{  }@{  }r|r@{  }@{  }r@{  }@{  }r@{  }l@{}|l@{ }|}
& &\multicolumn{3}{c|}{\textbf{Connectivity}} &\multicolumn{3}{c}{\textbf{LE-Lists}}&\multicolumn{1}{l@{}}{}&\multicolumn{1}{@{ }l@{ }}{\multirow{1}{*}{{\textbf{New}}}} \\
& &\bf Ours & \multicolumn{1}{@{}c@{  }}{\bf DHS'21} &\textbf{Spd.} & \textbf{Ours} & \multicolumn{1}{@{}c@{  }}{\textbf{Parlay}} & \textbf{Spd.} &\multicolumn{1}{l}{}&\multicolumn{1}{@{ }l@{ }}{\textbf{Graphs}} \\\cmidrule{1-8}
\parbox[t]{\tableside}{\multirow{4}{*}{\rotatebox[origin=c]{90}{\bf Social}}}
&OK &0.010 &0.008 &0.76 &0.577 &1.52 &2.63 &\multicolumn{1}{l}{}&\multicolumn{1}{@{}l}{\textbf{OK}~\cite{yang2015defining}}\\\cline{10-10}
&LJ &0.013 &0.012 &0.91 &0.502 &1.96 &3.91 &&com-orkut\\
&TW &0.093 &0.099 &1.05 &4.88 &16.6 &3.41 &&$n=3M$\\
&FT &0.197 &0.150 &0.76 &24.9 &30.0 &1.20 &&$m=234M$\\ \cline{10-10}
\cmidrule{1-8}
\parbox[t]{\tableside}{\multirow{4}{*}{\rotatebox[origin=c]{90}{\bf Web}}}
&SD &0.222 &0.271 &1.22 &13.9 &49.3 &3.56 &\multicolumn{1}{l}{}&\multicolumn{1}{@{}l}{\textbf{FT}~\cite{yang2015defining}}\\\cline{10-10}
&CW &2.425 &2.844 &1.17 &\multicolumn{3}{c}{out of memory} &&Friendster\\
&HL14 &3.694 &4.463 &1.21 &\multicolumn{3}{c}{out of memory} &&$n=65M$\\
&HL12 &8.446 &10.39 &1.23 &\multicolumn{3}{c}{out of memory} &&$m=3.6B$\\\cline{10-10}
\cmidrule{1-8}
\parbox[t]{\tableside}{\multirow{8}{*}{\rotatebox[origin=c]{90}{\bf KNN}}}
&HH5 &0.017 &0.041 &2.40 &2.06 &15.5 &7.55 &\multicolumn{1}{l}{}&\multicolumn{1}{@{}l}{\textbf{USA}~\cite{roadgraph}}\\\cline{10-10}
&CH5 &0.035 &0.026 &0.76 &5.38 &54.2 &10.1 &&RoadUSA\\
&GL2 &0.045 &0.132 &2.97 &2.89 &14.1 &4.87 &&$n=24M$\\
&GL5 &0.074 &0.177 &2.40 &12.4 &68.1 &5.49 &&$m=58M$\\
\cline{10-10}
&GL10 &0.123 &0.210 &1.71 &11.0 &56.6 &5.15 &\multicolumn{1}{l}{}&\multicolumn{1}{@{}l}{\textbf{GE}~\cite{roadgraph}}\\\cline{10-10}
\cline{10-10}
&GL15 &0.147 &0.236 &1.61 &11.7 &59.2 &5.06 &&Germany\\
&GL20 &0.166 &0.242 &1.46 &11.9 &63.7 &5.37 &&$n=12M$\\
&COS5 &1.310 &2.697 &2.06 &132 &329 &2.49 &&$m=32M$ \\\cline{10-10}
\cmidrule{1-8}
\parbox[t]{\tableside}{\multirow{2}{*}{\rotatebox[origin=c]{90}{\bf Road}}}
&USA &0.045 &0.092 &2.07 &14.9 &101 &6.74 &\multicolumn{1}{l}{}&\multicolumn{1}{l}{}\\
&GE &0.040 &0.129 &3.23 &5.98 &32.4 &5.42 &\multicolumn{1}{l}{}&\multicolumn{1}{l}{}\\
\cmidrule{1-8}
\parbox[t]{\tableside}{\multirow{4}{*}{\rotatebox[origin=c]{90}{\bf Lattice}}}
&SQR &0.161 &0.290 &1.80 &45.4 &184 &4.05 \\
&REC &0.023 &0.037 &1.66 &7.28 &520$^?$ &71.4 \\
&SQR' &0.134 &0.275 &2.06 &46.8 &202$^?$ &4.32 \\
&REC' &0.021 &0.043 &2.08 &8.57 &648$^?$ &75.7 \\
\end{tabular}
\vspace{-.15in}
\caption
{\small\textbf{Running time (in seconds) of connectivity and LE-Lists implementations.}
DHS'21$=$the LDD-UF-JTB connectivity implementation in \connectit{}~\cite{dhulipala2020connectit}.
Parlay$=$the LE-lists implementation in ParlayLib~\cite{blelloch2020parlaylib}.
Spd.$=$Baseline\_time $/$ our\_time.
``?''$=$results different from our parallel and sequential version, and the running time may not be accurate.
\label{tab:cclelists}
}
\end{table}

\subsection{Experiments on Connectivity and LE-Lists}\label{sec:exp-cc-lelists}
\myparagraph{Experiments on Connectivity.}\label{sec:exp-cc}
We implement the LDD-UF-JTB algorithm for graph connectivity in \connectit{} using our parallel hash bags (\cref{sec:hashbag}) to maintain the frontiers and the local search optimization (\cref{sec:local}). Both optimizations are applied to the sparse rounds in LDD.
In \cref{tab:cclelists}, we compare our algorithm to the same algorithm in \connectit{}.

On social networks with low diameters, our algorithm is slightly slower than \connectit{}, but is generally comparable.
This is because most of the vertices are visited in the dense mode, which is implemented similarly in both algorithms. 
The slowdown in our algorithm on social networks seems to be that \VGC{} brings more work to the first several sparse rounds, which reduces the benefit of using dense modes.
For other graph instances where dense modes do not significantly dominate the cost, our algorithm generally performs well.
On web graphs, our code is 1.21$\times$ faster than \connectit{} on average.
On the large-diameter 
graphs, our implementation is 1.98$\times$ faster than \connectit{} on average.
Since parallel hash bags and \VGC{} only apply to sparse rounds, the speedup of ours compared to \connectit{} has a correlation with the diameter of the graph.
Note that LDD is guaranteed to finish in $O(\log n)$ rounds, as opposed to $O(D)$ for diameter $D$ in SCC.
Therefore, the improvement of our implementation over \connectit{} is not as significant as the improvement of our SCC over existing work.
However, our implementation still outperforms \connectit{} on 16 instances out of 20, and is 1.67$\times$ faster than \connectit{} on average.
We believe that the experiments on connectivity provide additional evidence to show that our hash bags and \VGC{} are general and practical.

\hide{
We follow the same framework as \connectit{} in our implementation. We use LDD as our sampling phase and use union-find as our finish phase. We know it is impossible to compare against all implementations in \connectit{}. Thus, we compare our performance with \connectit{} using the same sampling and finish schemes. We do realize it is not the fastest implementation across all graphs, but it generally performs good.

\xiaojun{Should consider use \connectit{} or \gbbs{} to refer to their connectivity implementations}

Our connectivity implementation achieves 1.67$\times$ speedup over GBBS~\cite{gbbs} on average.
In particular, on large-diameter graphs, our implementation achieves 1.98$\times$ speedup over GBBS on average.
Our experiments show that our implementation outperforms GBBS on 20 instances out of 24.
We use almost the same framework as the compared implementation in GBBS.
Specifically, we use LDD with union-find to find the connected components.
However, unlike GBBS, we plug in the local search and \hashbag techniques in the LDD step, and we believe this is where the improvements come from.

On social networks, we do realize that our performance is slower than GBBS on social networks.
These graphs are relatively small and have small diameters.
The parallelism is already very good and the synchronization cost is small.
Using local search will result in the exponential growth in the frontier size, which leads to some overhead of maintaining more vertices in sparse mode.
We also notice that the running time on these graphs is negligible, so using either implementations will not make a huge difference.

\xiaojun{We need to think about how to explain our performance on CH5.}

Nevertheless, on large-diameter graphs, our implementation outperforms GBBS on almost all graphs. Note that connectivity is a well-studied problem. It is hard to gain improvements over the state-of-the-art graph processing system. Our experiments convince us that the local queue optimization and \hashbag is very efficient in practice.
} 

\myparagraph{Experiments on LE-lists.}
We compare our LE-lists implementation with ParlayLib~\cite{blelloch2020parlaylib} in \cref{tab:cclelists}.
Their implementation is the state-of-the-art and released in 2022.
Note that, unlike CC and SCC, here we can only use parallel hash bags for LE-lists but not \VGC{} since the BFS traverse orders need to be preserved.

On low-diameter graphs, our LE-list algorithm is 1.20--3.91$\times$ faster (2.73$\times$ on average) than \parlay{}'s implementation.
On large-diameter graphs, the speedup increases to 2.49--10.1$\times$ (5.36$\times$ on average).
We believe this is because \hashbag{s} maintain frontier more efficiently, and processing large-diameter graphs involve more rounds (frontiers).
Both our and \parlay{}'s implementation are unable to compute the LE-lists of the three largest graphs CW, HL14, and HL12, because the output size of LE-lists is $O(n\log n)$, which
is larger than the memory of our machine.
We also report the size of the LE-lists on each graph, and compare it to both \parlay{}'s implementation and Cohen's sequential algorithm~\cite{cohen1997size}.
\parlay{}'s implementation does not report the correct numbers on REC, SQR', and REC', and this is probably why they have poor performance on these graphs.

Overall, our algorithm is faster than \parlay{}'s implementation (the state-of-the-art implementation) on all graphs.
On average, our version is 4.34$\times$ faster than \parlay{}'s implementation on graphs with correct answers.
We note that it remains an interesting question on how to apply
a similar local search to LE-lists. We plan to study it in future work.

\hide{
Our parallel algorithm processes all the rest of the graphs in about two minutes, compared to up to one hour by the standard sequential algorithm.
The lowest speedup is on CH5 (2.7x) because CH5 is small but has a large diameter (even compared to other large-diameter graphs).
}

\hide{
Social graphs represent users and friendship on social networks.
Web graphs represent webpages or hosts and their hyperlinks.
Road networks represents roads and their connections on maps.
$k$-NN graphs are widely used in machine learning, such as graph clustering~\cite{Maier2009,Franti2006,Lucinska2012,karypis1999chameleon}, manifold learning~\cite{Tenenbaum2000}, outlier detection~\cite{Hautamaki2004}, and proximity search~\cite{Paredes2005,Chavez2010,Sebastian2002}.
Each node is multi-dimensional data point from machine learning datasets~\cite{}, and has $k$ edges to its $k$-nearest neighbors (excluding itself).
For synthetic graphs, we create four lattice graphs and two chains with different lengths. 
When creating 2D lattices (SQR and REC),
we add directed edges from $u$ to each of its neighbor $v$ with probability 0.5,
and from $v$ to $u$ with the same probability. We then remove duplicate edges.
We also created sampled lattice graphs (SQR' and REC'),
where we add directed edges from $u$ to each of its neighbors $v$ with probability $p=0.3$, and from $v$ to $u$ with probability $0.3$.
}

\section{Related Work}\label{sec:related}

Parallel SCC has been widely studied.
Prior to the BGSS algorithm based on (multi-)reachability searches, there had been other approaches.
The first type of approach is based on parallelizing DFS~\cite{bloemen2015fly,bloemen2016multi,lowe2016concurrent}.
However, since DFS is inherently sequential~\cite{reif1985depth} and hard to be parallelized,
these algorithms are shown to be slower than existing reachability-based solutions ~\cite{ji2018ispan}.
Another widely-adopted approach is based on single-reachability search (aka.\ the \emph{forward-backward search}, or \emph{Fw-Bw})~\cite{mclendon2005finding,coppersmith2003divide,fleischer2000identifying,hong2013fast,slota2014bfs,xu2018finding,ji2018ispan}.
However, \emph{Fw-Bw} does not provide sufficient parallelism to find all SCCs.
Hence, these systems only use \emph{Fw-Bw} to find large SCCs
and use other techniques such as coloring and trimming to find small SCCs, which do not have good theoretical guarantees.
For this type of approach, we compared the two newest ones with the released code: \multistep{}~\cite{slota2014bfs} and \ispan{}~\cite{ji2018ispan}.
They perform well on graphs with a small diameter and a large $\mathit{SCC}_1$ ($\mathit{SCC}_1$ is the largest SCC in the graph),
but do not work well on graphs with a large diameter or a small $\mathit{SCC}_1$ (e.g., the \knn{} and lattice graphs in our tests).

\hide{
Parallel SCC has also been studied on other platforms such as GPUs and distributed systems.
Comparing the wall-clock running times reported in the papers, it seems that shared-memory algorithms are much faster, but we note that different platforms have their own use cases.
On GPUs, many implementations are proposed (e.g.,~\cite{hou2020parallel,li2014efficient,barnat2011computing,stuhl2013computing,wijs2014gpu,devshatwar2016gpu,li2017high,hou2019parallel}).
Because of the limited size of GPU memory, we are not aware of existing GPU algorithms tested on graphs with more than 1 billion edges.
For some large graphs used in this paper, processing them on GPUs seems difficult.

Parallel SCC algorithms were also studied in the distributed settings (e.g.,~\cite{mclendon2005finding,malewicz2010pregel,salihoglu2014optimizing,yan2014pregel,barnat2011distributed}).
Most of them also use single-reachability searches and coloring, and Yan et al.'s algorithm~\cite{yan2014pregel} uses multi-reachability searches.
Although we tested the largest publicly-available graph (HL12) in this paper, we acknowledge that some private graphs (e.g., Google's graph and brain connectome) are larger and need to be processed on distributed systems.
}

Parallel SCC has also been studied on other platforms such as GPUs~\cite{hou2020parallel,li2014efficient,barnat2011computing,stuhl2013computing,wijs2014gpu,devshatwar2016gpu,li2017high,hou2019parallel} and distributed systems~\cite{mclendon2005finding,malewicz2010pregel,salihoglu2014optimizing,yan2014pregel,barnat2011distributed}.
Comparing the wall-clock running times reported in the papers, it seems that shared-memory algorithms are much faster, but we note that different platforms have their own use cases.

\myparagraph{Related Work of Parallel Hash Bag.} There exist other variants of hash tables designed for parallel algorithms~\cite{dong2021efficient,gu18algorithmic,Leiserson10}.
The \emph{parallel bag}~\cite{Leiserson10} supports similar interfaces as our \hashbag{}, but uses a very different design.
Parallel bags are organized using pointers, causing additional cache misses in practice.
Our \hashbag{} uses flat arrays and is practical and I/O-friendly.
The $k$-level hash table designed for NVRAMs~\cite{gu18algorithmic} requires allocating memory when resizing, while one of the goals of \hashbag{s} is to avoid explicit resizing.
Our work is also the first to formalize the interface of maintaining frontiers in parallel reachability search
and proposes a practical data structure (the \hashbag{}) with theoretical analysis.

\myparagraph{Parallel BFS.} There exist other implementations of parallel BFS, and some of them also consider reducing synchronization costs~\cite{beamer2015gap,zhang2018graphit,nguyen2013lightweight}. However, these implementations only consider a single source, and we are unaware of how to directly apply to the multi-reachability search needed in SCC.

\hide{
\myparagraph{DFS based implementations.}  DFS is considered hard to parallilize~\cite{reif1985depth}. Some novel parallel randomized DFS algorithms \cite{bloemen2015fly,bloemen2016multi} show DFS-based algorithms can be parallelized more directly, which have $O(p(n+m))$ work, where $p$ is the number of processors. Low and Gavin~\cite{lowe2016concurrent} proposed a concurrent DFS algorithm based on Tarjan's algorithm, which has $O(n^2)$ work. As shown in ~\cite{ji2018ispan}, \ispan{}~\cite{ji2018ispan} is more than 16 times faster than UFSCC~\cite{bloemen2016multi}, so we do not compare our algorithm with these DFS based implementations but with \ispan{}~\cite{ji2018ispan}.

\myparagraph{Reachability based implementations.} Most parallel SCC implementations use the idea of single-reachability (refered as Forward-Backward search in some papers) proposed in~\cite{coppersmith2003divide,fleischer2000identifying}. Usually, they use single-reachability search to find large SCCs, and use other algorithms, such as coloring and trimming~\cite{mclendon2005finding}, to deal with small SCCs.

For multi-core machines, Hone et al.~\cite{hong2013fast} explore the small-world property (low-diameter) for social networks and use recursive trimming and coloring algorithms to deal with small SCCs, which inspire many other implementations. Xu et al~\cite{xu2018finding} design parallel SCC algorithm based on generalized rough sets theory, on four-core machine, it performs similar to ~\cite{hong2013fast}. \multistep{}~\cite{slota2014bfs} also explore the low-diameter property, and gain better performance than ~\cite{hong2013fast}. So we do not compare with Hong~\cite{hong2013fast} in our paper but with \multistep{}~\cite{slota2014bfs}.

For distributed machines, Mclendon Iii et al.~\cite{mclendon2005finding} implemented single-reachability search algorithms on distributed systems and proposed the trimming technique.  Barnat J et al.~\cite{barnat2011distributed} exploy OWCTY elimination and use FW-BW and coloring algorithms, known as OBF algorithm.  Google's Pregel~\cite{malewicz2010pregel} proposed the vertice-centric computing paradigm for distributed systems. Two Pregel SCC implementations are proposed~\cite{salihoglu2014optimizing,yan2014pregel}. Note that Yan et al~\cite{yan2014pregel} proposed a multi-labeling algorithm, which is similar to the multi-reachability search in our paper, but multi-labeling is designed to increase the probability finding the largest SCC, and does not good theoretical work guarantee.

There are also lots of parallel implementations on GPUs~\cite{barnat2011computing,stuhl2013computing,wijs2014gpu,devshatwar2016gpu,li2017high,hou2019parallel}, which are able to get speedup on synthetic graphs. Due to the limitation of VRAM, the largest graph tested on GPUs is \emph{LiveJournal} (LJ)~\cite{backstrom2006group}. Besides implementations on common architectures, there is also an implementation for external memory systems~\cite{lv2016finding}.
}

\section{Discussions and Future Work}
\label{sec:discussions}

In this paper, we show that using faster algorithms on reachability queries can significantly accelerate the performance of SCC and related algorithms, especially for large-diameter graphs.
We tested our SCC algorithm on large-scale graphs with up to hundreds of billions of edges.
On average, our SCC algorithm is 6.0$\times$ faster than the best previous parallel implementation (\gbbs), 8.1$\times$ faster than \multistep, and 12.9$\times$ faster than Tarjan's sequential algorithms.

\revision{
We believe that the two key techniques in this paper, the hash bag and vertical granularity control, are general and of independent interest.
In this paper, we apply them to graph connectivity and LE-lists.
The experimental results show that they lead to improved performance than prior work.
We believe that they also apply to many other applications.

Hash bags are used to maintain frontiers (a subset of vertices) in graph algorithms.
Many state-of-the-art graph libraries (e.g., \gbbs~\cite{gbbs2021} and Ligra~\cite{shun2013ligra}) use the abstract data type (ADT) called VertexSubset to maintain frontiers
on many graph algorithms.
Hash bags can be used to implement this ADT by replacing the current data structure (fixed-size array). 
With careful engineering, we believe hash bags can potentially improve the performance of these implementations. We leave this as future work.

The high-level idea of \VGC{} applies to traversal-based graph algorithms, 
such as BFS, algorithms for connectivity, biconnectity, single source shortest paths (SSSP), and some others in~\cite{dhulipala2017,gbbs2021,shun2013ligra}.
\VGC{} can potentially accelerate them on large-diameter graphs.
Our specific ``local-search'' idea does not directly apply as is. 
When the traversing order does not matter (e,g., reachability-based algorithms), local search can be applied directly. 
In a recent paper, we apply local search to graph biconnectivity~\cite{bccanounymous}, which improved the overall performance by up to 4$\times$ on a variety of graphs.
For some distance-based algorithms, we need additional designs on top of local-search,
such as supporting revisiting certain vertices (e.g., in BFS, SSSP, LE-lists) for relaxation,
or some wake-up strategies to find the next frontier (e.g., in $k$-core).
We believe that this is an interesting research direction, and plan to explore it in the future.
} 

\section*{Acknowledgement}
This work is supported by NSF grants CCF-2103483, CCF-2238358, IIS-2227669, and UCR Regents Faculty Fellowships.
We thank anonymous reviewers for the useful feedbacks. 

\appendix
 \section{Proof of \cref{thm:hashbag}}
\label{app:hashbagsize}

\medskip
\begin{proof} (sketch) 
  We first show the algorithm is correct (i.e., each chuck will not be full), and then analyze the cost bounds.
  We consider the sampling in insertions as a coin-tossing process.
  Let $k$ be the chunk size, and the sample rate is $\ssize/\alpha k$. 
  We first consider the sequential case ($P=1$). We are interested in the number of insertions performed before the number of samples reaches $\ssize$. 
  This is equivalent to the number of coins we toss before we see $\ssize$ heads (sampled insertions). 
  This can be analyzed using the Chernoff bound---assuming the $\ssize=\Theta(\log n)$ and $k=\Omega(\ssize/\epsilon^2)$,  the number of tossed coins $e$ satisfies $(1-\epsilon)\alpha k\le e\le (1+\epsilon)\alpha k$ for $0<\epsilon<1$ \whp.
  If we set $\alpha=\epsilon=0.5$, the load factor for this chunk when resizing is between $1/4$ and $3/4$ \whp. 

  Now consider the parallel case, where $P$ processors execute insertions asynchronously.
  In addition, when an insertion sampled, it first increments the counter using \cas{}, which can be delayed by other processors. 
  Based on the assumption given in \cref{sec:hashbag}, between two consecutive \cas{s}, each of the other processors can only execute a constant number of instructions, so a total of $O(P)$ insertions can be done.
  Since \cas{} is atomic, one processor has to win and proceed, so an unsuccessful \cas{} can happen for at most $\ssize$ times before resizing. 
  As compared to the sequential setting, $O(P\ssize)$ more elements can be inserted. 
  When $k=\Omega(P\ssize+\ssize/\epsilon^2)$, each chunk of the hash bag will not be overfull \whp{}.

  We now show that for $s$ insertions, all elements in the \hashbag will be in the first $O(s+\lambda)$ positions in the $\bagarray$ array \whp{}, which bounds the work and span for listing and packing.
  As discussed above, the ``wasted'' space due to parallelism is upper bounded by $O(P\ssize)$, which is asymptotically bounded even for the first chunk with $\lambda=\Omega(\ssize(P+\log n))$.
  Since the load factor for each chunk is a constant when resizing, the total size of all chunks in use is $O(s+\lambda)$.

  Finally, we show the costs for insertions.
  An insertion to a hash table with linear-probing and a constant load factor uses $O(1)$ expected work and $O(\log n)$ span \whp{}.
  For updating the sample count, there can be at most $O(\log s\log n)$ samples \whp{}, which is also the longest possible dependence chain in the algorithm.
  If we assume $\lambda=\Omega(\ssize(P+\log n))$, for each insertion, the probability of it to be picked as a first chunk sample is $O(1/(P+\log n))=O(1/\log n)$, and in the worst case we need to wait for $\ssize=\Theta(\log n)$ retries until the counter is incremented.
  Hence, the amortized work is $O(1)$.
  Then the sample rate halves for each of the next chunks, so taking the sum of a geometric series, the total amortized work to maintain counters is $O(1)$ for an insertion.
\end{proof}

Guided by the theory, we set $\lambda=2^{14}$ and $\ssize=50$.
We note that if the frontier size is smaller than $\lambda$, this may cause $O(\textsf{diam}(G)\cdot\lambda)$ work (which can be more than $m$) for a reachability search, since the work in each round is $\Omega(\lambda)$. 
To maintain the work bound of reachability searches, we want to guarantee the frontier size is $\Omega(\lambda)$ except for possibly the last frontier. This can be guaranteed by setting the local queue size (see \cref{sec:local}) to be $\Omega(\lambda)$.

\hide{
  We argue that we will resize before we have $P+q$ samples.
  Consider the worst case that if a processor keeps executing the non-sampled insertions, and stops if it sees a sampled insertion.
  We will see $P$ samples.
  Then some processors start to \faa{} using \cas{}.
  Since \cas{} is atomic, one processor has to win and proceed, so this process can only repeat for $q$ times before resizing.
  In total, a resizing is triggered before we see $P+q$ samples.
  Again consider the coin-tossing process, and we want to consider the number of coin tosses before seeing $P+q$ heads.
  The same as above, by setting $\ssize=\Theta(\log n)$ and $k\ge \lambda=\Omega((P+\log n)(1/\epsilon^2+\log n))$, the Chernoff bound shows the number of tossed coins is no more than $(1+\epsilon)\alpha k$ \whp.
}




\bibliography{main}

\clearpage

\end{document}